\documentclass[prd,twocolumn]{revtex4}
\usepackage{dcolumn}
\usepackage{multirow}
\usepackage{amsmath,amsthm,amssymb}
\usepackage{hyperref}

\def\fsc{A}
\def\lsc{\mathpzc{l}}
\def\Asc{\mathpzc{A}}
\def\Bsc{\mathpzc{B}}

\def\Nsc{\mathpzc{N}}
\def\Psc{\mathpzc{P}}

\newcommand{\pert}[2]{
{}^{(#1)}\hspace{-0.5mm}#2
}

\DeclareMathAlphabet{\mathpzc}{OT1}{pzc}{m}{it}

\begin{document}
%


\preprint{CERN-TH-2017-023}
\title{Higher order perturbations of Anti-de Sitter space and time-periodic solutions of vacuum Einstein equations}

\author{Andrzej Rostworowski}
\email{arostwor@th.if.uj.edu.pl}
\affiliation{M. Smoluchowski Institute of Physics, Jagiellonian University, 30-348 Krak\'ow, Poland}
\affiliation{Theoretical Physics Department, CERN, CH-1211 Geneva 23, Switzerland}
%
%
\begin{abstract}
Motivated by the problem of stability of Anti-de Sitter (AdS)
spacetime, we discuss nonlinear gravitational perturbations of
maximally symmetric solutions of vacuum Einstein equations in general
and the case of AdS in particular. We present the evidence that,
similarly to the self-gravitating scalar field at spherical symmetry,
the negative cosmological constant allows for the existence of globally
regular asymptotically AdS, time-periodic solutions of vacuum Einstein
equations whose frequencies bifurcate from linear eigenfrequencies of
AdS. Interestingly, our preliminary results indicate that the number
of one parameter families of time-periodic solutions bifurcating from
a given eigenfrequency equals the multiplicity of this eigenfrequency.
\end{abstract}

\maketitle

\section{Introduction.}
The problem of late-time dynamics in asymptotically Anti-de Sitter
(AdS) spacetimes received quite a lot of attention in the past five
years and revealed a few surprises. First, the paper \cite{br}
provided numerical and heuristic evidence for two types of possible
scenarios in the model Einstein-AdS--massless scalar field system at
spherical symmetry: 1) turbulent dynamics leading to concentration of
(a fraction of) energy on small spatial scales leading to a black hole
formation on the timescale $\mathcal{O}\left( \epsilon^{-2}\right)$,
where $\epsilon$ measures the amplitude of initial data, and 2)
quasiperiodic evolution. 
Second, stable time-periodic solutions for this model were discovered
\cite{mrPRL}, providing an explanation for quasiperiodic evolution, as
being anchored to stability islands of these time periodic solutions
(see also \cite{mr@bll} vs. \cite{bll_BS}) and an important role of
nondispersive spectrum of linear perturbations for instability was
identified \cite{dhms,mrPRD}. Finally a new perturbation scheme was
proposed \cite{bbgll} and firmly developed \cite{cev1,cev2} to capture the
dynamics on the $\mathcal{O}\left( \epsilon^{-2}\right)$ timescale. It consists
in setting time-averaged system of equations, for the slow time
dependence of Fourier-like coefficients ("slow" with respect to "fast"
time-scale of Fourier modes themselves). The main advantage of this
system (hereafter referred to as the resonant system) is that it has a
scaling symmetry: if its solution with the initial amplitude 1 does
something at time $t$, then the corresponding solution with the
initial amplitude $\epsilon$ does the same thing at time
$t/\epsilon^2$. Thus, by solving the resonant system one can probe the
regime of arbitrarily small perturbations (whose outcome of evolution
of full Einstein equations is beyond the possibility of numerical
verification). Interestingly, the solutions of the resonant system
generically develop an oscillatory blow-up in finite time giving an
extra piece of evidence for instability of AdS \cite{bmr}.

After gaining some insights on AdS stability problem from a toy model
of a matter field coupled to Einstein equations at spherical symmetry,
it becomes crucial to decide if it constitutes a good model for a
general setting and for Einstein equations in vacuum in
particular. Extrapolating from \cite{br,mrPRL} beyond spherical
symmetry one could make the following two conjectures (for effectively
reflecting boundary conditions at timelike boundary of AdS): 1) the
turbulent instability is present in AdS i.e. the transfer of energy to
arbitrarily high frequencies does not saturates and probably leads to
a gravitational collapse, 2) there exist globally regular,
time-periodic, asymptotically AdS solutions of Einstein equations
(\textit{TP solutions}, also refereed to as \textit{geons} in
\cite{hs,dhs,ds}), immune to this instability; their frequencies are
expected to bifurcate from eigenfrequencies of the spectrum of linear
perturbations of AdS. One possible way to build evidence for these
conjectures, is to set the resonant system for vacuum Einstein
equations (to study instability) and to construct TP solutions
perturbatively. To achieve these goals one has to master gravitational
perturbations of AdS (in principle up to arbitrarily high orders to
construct TP solutions, but only up to the third order to construct
the resonant system).

There is a limited number of results on the AdS instability outside spherical symmetry. Horowitz and Santos showed numerically
that in the vacuum case there are linear modes that can be extended to
form TP solutions \cite{hs}. The results of perturbative calculations
performed around some single modes (or some linear combination of
modes) up to the third order were given in \cite{dhs} and its sequel
\cite{ds}. However a systematic approach to gravitational
perturbations of AdS has never been presented. 
The main aim of this work is to fill in this gap and to treat higher
order perturbations of AdS in a systematic manner. We hope that our
results will provide a solid base for the future construction of the
resonant system for vacuum AdS and for more systematic studies of
regular, asymptotically AdS, time-periodic solutions of Einstein
vacuum equations. 
We restrict ourselves to the case of axially symmetric
perturbations as a relatively simple and illustrative example. All
conceptual difficulties are encountered when stepping out from
spherical symmetry to axial symmetry. Once this model case is well
understood, adding azimuthal angle dependence (in 3+1 dimensions) is a
technical, not a conceptual, issue.

The novel feature of vacuum perturbation with respect to spherically
symmetric selfgravitating scalar field model is the degeneracy in the
spectrum of linear perturbations. Namely, the spectrum of linear
perturbation in $3+1$ dimensions reads $\omega_{\ell,j} = q+\ell+2j$,
where $\ell$ is the mode angular momentum index, $j$ is the nodal
number of the corresponding radial wave function and $q$ reads $1$ or
$2$ for polar and axial modes respectively. It was noted with a bit of
surprise that only in special cases (listed in Sec.6 in \cite{ds})
linear eigenmodes admit a nonlinear extension to a TP
solution. However, it is well known since quantum mechanics that the
degeneracy case may require a special treatment in perturbation
expansion, and a more appropriate question to ask is which linear
frequencies (rather then which linear modes) are bifurcation points
for TP solutions. It seems that, similarly to spherically symmetric
scalar field case, there are one parameter families of TP solutions,
bifurcating from \textit{each} linear frequency and the number of
these families of TP solutions is \textit{equal} to the multiplicity
of the linear eigenfrequency \cite{r}. In fact, the key property
allowing the time-periodic solutions of the type studied in
\cite{mrPRL} (i.e. bifurcating from a single linear eigenfrequency) to
exist, is the absence of $(+++)$ resonances, see \cite{cev1} for the
definition and a brute force proof for the Einstein-AdS-massless
scalar field system (for a deeper discussion and the proof of this
property for a test scalar field on AdS background see \cite{Evnin}
and references therein). It seems that there are no $(+++)$ resonances
also in vacuum AdS case and, in this sense, the self-gravitating
scalar field seems to be a good toy model to study gravitational
stability of AdS.

We concentrate in this work on (nonlinear) perturbations of AdS space,
however it is quite clear that similar approach can be taken for
perturbations of any spherically symmetric solution of vacuum Einstein
equations. In fact all formulae in the paper, that are not specific to
a particular asymptotic structure of the spacetime are general enough
to work for any maximally symmetric solution of vacuum Einstein
equations (Anti-de Sitter, Minkowski and de Sitter) in static
coordinates, i.e. with the line element in the form
(\ref{eq:line_element}). The key observation is to split perturbative
Einstein equations into these satisfied identically by a suitable
choice of master scalar variables (see below) and those that govern
the dynamics of these master scalar variables. To the best of our
knowledge this splitting has never been highlighted in the literature.

The paper is organized as follows. In Sec.~2 we discuss the general
setup for perturbation expansion around exact solutions of vacuum
Einstein equations, and Regge-Wheeler \cite{RW} decomposition of
metric perturbations. In Secs.~3~and~4 we discuss polar and axial type
perturbations of maximally symmetric spacetimes. We apply the
formalism of these sections in Sec.~5 to construct and study some
properties of globally regular, aAdS, time-periodic solutions of
vacuum Einstein equations. Then we conclude in Sec.~6.

\begin{widetext}
\section{$AdS_{d+1}$ perturbation in vacuum - general setup}

We are interested in solutions of vacuum Einstein equations with
negative cosmological constant $\Lambda= - \frac{d(d-1)}{2 \lsc^2}$:
\begin{equation}
\label{Eeqs}
R_{\mu \nu} + \frac{d}{\lsc^2} g_{\mu \nu} = 0.
\end{equation} 
Let the "bar" quantities stand for the AdS quantities (i.e. $\bar g_{\mu\nu}$, $\bar g^{\mu\nu}$, $\bar \Gamma^{\alpha}_{\mu \nu}$, $\bar R_{\mu\nu}$ are AdS quantities). Now let $g_{\mu\nu} = \bar g_{\mu \nu} + \delta g_{\mu\nu}$ or in matrix notation $g=\bar g + \delta g$, where $\delta g_{\mu\nu}$ will be expanded as 
\begin{equation}
\label{pertSer}
\delta g_{\mu\nu} = \sum_{1 \leq i} \pert{i}{h}_{\mu \nu} \, \epsilon^i
\end{equation}
later on. Then we have:
\begin{eqnarray}
\label{eq:gInverse}
g^{\alpha \beta} &=& \left(\bar g^{-1} - \bar g^{-1} \delta g \bar g^{-1} + \bar g ^{-1} \delta g \bar g ^{-1} \delta g \bar g ^{-1} - \dots \right)^{\alpha \beta} 
\nonumber\\
&=& \bar g^{\alpha \beta} + \delta g^{\alpha \beta} \,,
\\
\label{eq:GammaPert}
\Gamma^{\alpha}_{\mu \nu} &=& \bar \Gamma^{\alpha}_{\mu \nu} + \frac {1} {2} \left(\bar g^{-1} - \bar g^{-1} \delta g \bar g^{-1} + \bar g ^{-1} \delta g \bar g ^{-1} \delta g \bar g ^{-1} - \dots \right)^{\alpha \lambda} (\bar \nabla_{\mu} \delta g_{\lambda \nu} + \bar \nabla_{\nu} \delta g_{\lambda \mu} - \bar \nabla_{\lambda} \delta g_{\mu \nu}) 
\nonumber\\
&=& \bar \Gamma^{\alpha}_{\mu \nu} + \delta \Gamma^{\alpha}_{\mu \nu} \, ,
\\
\label{eq:RPert}
R_{\mu \nu} &=& \bar R_{\mu \nu} + \bar \nabla_{\alpha} \delta \Gamma^{\alpha}_{\mu \nu} - \bar \nabla_{\nu} \delta \Gamma^{\alpha}_{\alpha \mu} + \delta \Gamma^{\alpha}_{\alpha \lambda} \delta \Gamma^{\lambda}_{\mu \nu} - \delta \Gamma^{\lambda}_{\mu \alpha} \delta \Gamma^{\alpha}_{\lambda \nu} 
\nonumber\\
&=&  \bar R_{\mu \nu} +  \delta R_{\mu \nu} \, .
\end{eqnarray} 
The equation (\ref{eq:gInverse}) is straightforwardly obtained by recursive application of the formula $g^{\alpha \beta} = \bar g^{\alpha \beta} - \bar g^{\alpha \mu} \delta g_{\mu \nu} g^{\nu \beta}$ and then the eqs. (\ref{eq:GammaPert},\ref{eq:RPert}) easily follow. Thus the Einstein equations read ("bar" quantities are solutions to Einstein equations)
\begin{equation}
\label{EeqsPert}
\delta R_{\mu \nu} + \frac{d}{\lsc^2} \delta g_{\mu \nu} = 0.
\end{equation} 
Now, plugging the expansion (\ref{pertSer}) into (\ref{EeqsPert}) and collecting the terms at the same powers of $\epsilon$, we get the following hierarchy of equations:
\begin{equation}
\label{eq:pertEq}
\pert{i}{E}_{\mu\nu} := \Delta_L \pert{i}{h}_{\mu \nu} - \pert{i}{S}_{\mu \nu} = 0\,.
\end{equation}
The \textit{Lorentzian} Lichnerowicz operator $\Delta_L$ reads
\begin{eqnarray}
\Delta_L h_{\mu \nu} &=& \frac {1} {2} \bar g^{\alpha \lambda} \left( \bar \nabla_{\alpha} (\bar \nabla_{\mu} h_{\lambda \nu} + \bar \nabla_{\nu} h_{\lambda \mu} - \bar \nabla_{\lambda} h_{\mu \nu}) - \bar \nabla_{\nu} (\bar \nabla_{\alpha} h_{\lambda \mu} + \bar \nabla_{\mu} h_{\lambda \alpha} - \bar \nabla_{\lambda} h_{\alpha \mu}) \right) + \frac{d}{\lsc^2} h_{\mu \nu} 
\nonumber\\
&=& \frac{1}{2} \left(- \bar \nabla^{\alpha} \bar \nabla_{\alpha} h_{\mu \nu} - \bar \nabla_{\mu} \bar \nabla_{\nu} h - 2 \bar R_{\mu \alpha \nu \beta} h^{\alpha \beta} + \bar \nabla_{\mu} \bar \nabla^{\alpha} h_{\nu \alpha} + \bar \nabla_{\nu} \bar \nabla^{\alpha} h_{\mu \alpha} \right) \,,  
\label{eq:Delta_L}
\end{eqnarray}
where $h = \bar g^{\alpha \beta} h_{\alpha \beta}$, $h^{\alpha \beta} = \bar g^{\alpha \mu} \bar g^{\beta \nu} h_{\mu \nu}$ and we have used
\[
\left( \bar \nabla_{\alpha} \bar \nabla_{\nu} - \bar \nabla_{\nu} \bar \nabla_{\alpha} \right) h^{\alpha}_{\,\,\,\mu} = \bar R^{\alpha}_{\,\,\,\beta \alpha \nu} h^{\beta}_{\,\,\,\mu} - \bar R_{\beta \mu \alpha \nu} h^{\alpha \beta} = - \frac{d}{\lsc^2} h_{\nu \mu} - \bar R_{\beta \mu \alpha \nu} h^{\alpha \beta}  \,. 
\]
The source terms in (\ref{eq:pertEq}) read:
\begin{align}
\pert{i}{S}_{\mu \nu} = \left[\epsilon^i \right] & \left\{ 
- (1/2) \bar \nabla_{\alpha} \left[ \left(- \bar g^{-1} \delta g \bar g^{-1} + \bar g ^{-1} \delta g \bar g ^{-1} \delta g \bar g ^{-1} - \dots \right)^{\alpha \lambda} \left(\bar \nabla_{\mu} \delta g_{\lambda \nu} + \bar \nabla_{\nu} \delta g_{\lambda \mu} - \bar \nabla_{\lambda} \delta g_{\mu \nu}\right)\right] \right.
\nonumber\\
& \hskip 1.4mm + (1/2) \bar \nabla_{\nu} \left[\left(- \bar g^{-1} \delta g \bar g^{-1} + \bar g ^{-1} \delta g \bar g ^{-1} \delta g \bar g ^{-1} - \dots \right)^{\alpha \lambda} \left(\bar \nabla_{\mu} \delta g_{\lambda \alpha} + \bar \nabla_{\alpha} \delta g_{\lambda \mu} - \bar \nabla_{\lambda} \delta g_{\mu \alpha}\right)\right]
\nonumber\\
& \left. \hskip 1.4mm - \delta \Gamma^{\alpha}_{\alpha \lambda} \delta \Gamma^{\lambda}_{\mu \nu} + \delta \Gamma^{\lambda}_{\mu \alpha} \delta \Gamma^{\alpha}_{\lambda \nu} \right\} \,,
\label{eq:source_definition}
\end{align} 
where $\left[ \epsilon^i \right]f$ denotes the coefficient at
  $\epsilon^i$ in the (formal) power series expansion of $f=\sum_i f_i
  \epsilon^i$. The formulae
  (\ref{eq:pertEq}-\ref{eq:source_definition}) are in fact completely
  general i.e. they work for any value of cosmological constant and
  any zero order solution, no matter if spherically symmetric or
  not. From now on we limit to $d=3$ spatial dimensions.

Before presenting all further technicalities let us start with discussing
general strategy. We will follow the Regge-Wheeler (RW) seminal paper
\cite{RW} and use spherical symmetry of the zero order solution (AdS
in our case) to expand the metric perturbations into scalar, vector,
and tensor spherical harmonics. The reader unfamiliar with the
RW decomposition can consult Sec.~2 of the excellent review
by Nollert \cite{Nollert}. After separating angular dependence, the
system of ten perturbative Einstein equations (\ref{eq:pertEq}) splits
into the system of seven equations for polar (alternatively called
scalar) type perturbations and the system of three equations for axial
(alternatively called vector) type perturbations \cite{RW,
  Nollert}. The polar and axial parts decouple at linear
order, but generally mix at higher orders. The axial symmetry is
exceptional in this respect, as if we excite only polar perturbations
at linear order we stay in polar sector at all higher orders as well,
while starting with only axial perturbations at linear order results in axial
perturbations at all odd orders and polar perturbations at all even
orders of perturbation expansion. The most general gauge vector (for
given spherical harmonics indexes $\ell,m$) splits accordingly into
polar and axial parts parametrized by three and one function,
respectively. It turns out that the Lichnerowicz operator
(\ref{eq:Delta_L}) depends only on four and two RW gauge invariant
variables in polar and axial sector respectively, while the sources
(\ref{eq:source_definition}) contain also gauge degrees of
freedom. After introducing RW variables the system is still messy,
however it is well known that at linear order both polar and axial
sector are governed by only one (for given $\ell$, $m$) master scalar
variable each. It is very convenient to introduce corresponding scalar
variables also at higher orders - not only to ease the solution of the
system of Einstein equations but also to make the resonant structure
of these equations explicit. There are however a few problems to be
faced here:
\begin{enumerate}
\item 
to find the correct definition of master scalar variables at higher
orders,
\item 
to build the source term for the (inhomogeneous) wave equation for
this scalar variable from the sources (\ref{eq:source_definition}) in
perturbative Einstein equations (\ref{eq:pertEq}),
\item 
to be able to switch between the components of metric perturbations and
the master scalar variables (to solve Einstein equations easily) and back (to
reconstruct the metric perturbations from scalar variables and to be
able to find the sources (\ref{eq:source_definition}) necessary to
step to the next order of the perturbation expansion),
\item 
to set the metric perturbations to asymptotically AdS (aAdS) form with
a suitable gauge transformation (it is necessary to insure the correct
asymptotics of the sources to step to the next order),
\item 
to treat the special cases $\ell=0$ (for polar type perturbations only) and
$\ell=1$ (both for polar and axial type perturbations); they
correspond to gauge degrees of freedom at linear order but have to be
correctly dealt with at higher orders.
\end{enumerate}
In the two following sections we discuss solutions to all problems
listed above for axially symmetric perturbations starting from polar
and axial type perturbations at linear order. In what follows we use
coordinates $(t,r,\theta,\phi)$ in which the AdS line element reads
\begin{equation}
\label{eq:line_element}
ds^2 = -\fsc dt^2 + \fsc^{-1} dr^2 + r^2 d\Omega^2_2 \, ,
\end{equation}
with $\fsc \equiv \fsc(r) = \left(1 + r^2/\lsc^2\right)$. Then we use Mukohyama
\cite{Mukohyama}, appendix B definitions for scalar, vector and tensor
spherical harmonics (up to normalization factors). In $3+1$ dimensions
any tensor $T_{\mu\nu}$ can be split into seven polar and three axial
components. For any tensor $T_{\mu\nu}$ its polar components expanded
into (one scalar, one vector and two tensor) polar spherical harmonics
at axial symmetry read
\begin{align}
\label{eq:polarT_ab}
T_{ab}(t,r,\theta) = & \sum_{\ell} T_{\ell\,\,ab} (t,r) P_{\ell}(\cos\theta), \quad a,b=0,1 \, , 
\\
\label{eq:polarT_a2}
T_{a2}(t,r,\theta) = & \sum_{\ell} T_{\ell\,\,a2} (t,r) \partial_\theta P_{\ell}(\cos\theta), \quad a=0,1 \, , 
\\
\label{eq:polarT+}
\frac{1}{2}\left(T_{22}(t,r,\theta) + \frac{T_{33}(t,r,\theta)}{\sin^{2}\!\theta}\right) = & \sum_{\ell} T_{\ell\,\,+} (t,r) P_{\ell}(\cos\theta) \, ,
\\
\label{eq:polarT-}
\frac{1}{2}\left(T_{22}(t,r,\theta) - \frac{T_{33}(t,r,\theta)}{\sin^{2}\!\theta}\right) = & \sum_{\ell} T_{\ell\,\,-} (t,r) \left(-\ell(\ell+1) P_{\ell}(\cos\theta) - 2 \cot{\theta} \partial_{\theta}P_{\ell}(\cos\theta)\right) \, ,
\end{align}
where $P_{\ell}$ are Legendre polynomials. For any tensor $T_{\mu\nu}$ its axial components expanded into (one
vector and one tensor) axial spherical harmonics at axial symmetry
read
\begin{align}
\label{eq:axialT_a3}
T_{a3}(t,r,\theta) = & \sum_{\ell} T_{\ell\,\,a3} (t,r) \sin \theta \partial_{\theta} P_{\ell}(\cos\theta), \quad a=0,1 \, , 
\\
\label{eq:axialT_23}
T_{23}(t,r,\theta) = & \sum_{\ell} T_{\ell\,\,23} (t,r) \left(- 2 \cos \theta \partial_{\theta} P_\ell(\cos\theta) - \ell(\ell+1) \sin \theta P_\ell(\cos\theta) \right) \, .
\end{align}
Accordingly, in $3+1$ dimensions
any vector $V_{\mu}$ can be split into three polar and one axial
component. For any vector $V_{\mu}$ its polar components expanded
into (one scalar and one vector) polar spherical harmonics
at axial symmetry read
\begin{align}
\label{eq:polarV_a}
V_{a}(t,r,\theta) = & \sum_{\ell} V_{\ell\,\,a} (t,r) P_{\ell}(\cos\theta), \quad a = 0,1 \, , 
\\
\label{eq:polarV_2}
V_{2}(t,r,\theta) = & \sum_{\ell} V_{\ell\,\,2} (t,r) \partial_\theta P_{\ell}(\cos\theta) \, . 
\end{align}
For any vector $V_{\mu}$ its axial component expanded into (one
vector) axial spherical harmonics at axial symmetry
reads
\begin{align}
\label{eq:axialV_3}
V_{3}(t,r,\theta) = & \sum_{\ell} V_{\ell\,\,3} (t,r) \sin \theta \partial_{\theta} P_{\ell}(\cos\theta) \, . 
\end{align}
In what follows, the symbols $\pert{i}{h}_{\ell\,\,\mu\nu}$, $\pert{i}{S}_{\ell\,\,\mu\nu}$, $\pert{i}{E}_{\ell\,\,\mu\nu}$ and $\Delta_L \pert{i}{h}_{\ell\,\,\mu\nu}$ appear in expansion of \textit{tensors} $\pert{i}{h}_{\mu\nu}$, $\pert{i}{S}_{\mu\nu}$, $\pert{i}{E}_{\mu\nu}$ and $\Delta_L \pert{i}{h}_{\mu\nu}$ according to (\ref{eq:polarT_ab}-\ref{eq:axialT_23}) respectively (cf. (\ref{pertSer},\ref{eq:pertEq})). The symbols $\pert{i}{\zeta}_{\ell\,\,\mu}$ and $\pert{i}{\eta}_{\ell\,\,\mu}$ appear in expansion of polar gauge \textit{vectors} $\pert{i}{\zeta}_{\mu}$ and axial gauge \textit{vectors} $\pert{i}{\eta}_{\mu}$ according to (\ref{eq:polarV_a}-\ref{eq:axialV_3}) respectively (see below for the usage of these gauge vectors). 

\section{Polar perturbations at axial symmetry}
Although the polar type perturbations are technically much more involved
than axial type perturbations (in the case of linear Schwarzschild
perturbation it took thirteen years after solving axial linear
perturbations \cite{RW} to solve the polar ones \cite{zerilli}), we
start the discussion of gravitational perturbations with polar type
perturbations as they are indispensable at nonlinear level: even if we
start with axial type perturbations at linear order we end up with
polar type perturbations at second order of perturbation expansion. In
axial symmetry for polar perturbations we have
\begin{equation}
  \label{eq:hPolar}
  \left( \pert{i}{h}_{\alpha\beta} \right) = 
\left(\begin{array}{cccc}
    \pert{i}{h}_{00} & \pert{i}{h}_{01} & \pert{i}{h}_{02} & 0 \\
    \pert{i}{h}_{01} & \pert{i}{h}_{11} & \pert{i}{h}_{12} & 0 \\
    \pert{i}{h}_{02} & \pert{i}{h}_{12} & \pert{i}{h}_{22} & 0 \\
    0 & 0 & 0 & \pert{i}{h}_{33}  
\end{array}\right) \, ,
\end{equation}
with (cf. (\ref{eq:polarT_ab}-\ref{eq:polarT-}))
\begin{align}
\label{eq:h00}
\pert{i}{h}_{\ell\,\,00} = & \pert{i}{f}_{\ell\,\,00} + 2 \partial_t \pert{i}{\zeta}_{\ell\,\,0} - \fsc \, \fsc' \, \pert{i}{\zeta}_{\ell\,\,1} \, ,
\\
\pert{i}{h}_{\ell\,\,11} = & \pert{i}{f}_{\ell\,\,11} + 2 \partial_r \pert{i}{\zeta}_{\ell\,\,1} + \frac{\fsc'}{\fsc} \pert{i}{\zeta}_{\ell\,\,1} \, ,
\\
\pert{i}{h}_{\ell\,\,01} = & \pert{i}{f}_{\ell\,\,01} + \partial_r \pert{i}{\zeta}_{\ell\,\,0} + \partial_t \pert{i}{\zeta}_{\ell\,\,1} - \frac{\fsc'}{\fsc} \pert{i}{\zeta}_{\ell\,\,0} \, ,
\\
\pert{i}{h}_{\ell\,\,02} = & \pert{i}{\zeta}_{\ell\,\,0} + \partial_t \pert{i}{\zeta}_{\ell\,\,2} \, ,
\\
\pert{i}{h}_{\ell\,\,12} = & \pert{i}{\zeta}_{\ell\,\,1} - \frac{2}{r}\pert{i}{\zeta}_{\ell\,\,2} + \partial_r \pert{i}{\zeta}_{\ell\,\,2} \, ,
\\
\pert{i}{h}_{\ell\,\,+}  = & r^2 \pert{i}{f}_{\ell\,\,+} + 2 r \fsc \, \pert{i}{\zeta}_{\ell\,\,1} - \ell(\ell+1) \pert{i}{\zeta}_{\ell\,\,2} \, ,
\\
\label{eq:h-}
\pert{i}{h}_{\ell\,\,-} = & \pert{i}{\zeta}_{\ell\,\,2} \, , 
\end{align}
where $\pert{i}{\zeta}_{\ell\,\,0}$, $\pert{i}{\zeta}_{\ell\,\,1}$,
$\pert{i}{\zeta}_{\ell\,\,2}$ polar components define the $i$-th order
polar gauge vector $\pert{i}{\zeta}_{\mu}$
(cf. (\ref{eq:polarV_a},\ref{eq:polarV_2})) and
$\pert{i}{f}_{\ell\,\,00}(t,r)$, $\pert{i}{f}_{\ell\,\,11}(t,r)$,
$\pert{i}{f}_{\ell\,\,01}(t,r)$, $\pert{i}{f}_{\ell\,\,+}(t,r)$ are
Regge-Wheeler variables \cite{RW, Nollert}, being gauge invariant with
respect to gauge transformations induced by $\pert{j}{\zeta}_{\mu}$ with
$j \geq i$, i.e. gauge transformations of the form
\begin{equation}
\label{eq:gaugeTransform}
\sum_{1 \leq i} \pert{i}{h}_{\mu \nu} \, \epsilon^i \rightarrow \sum_{1 \leq i} \pert{i}{h}_{\mu \nu} \, \epsilon^i + \epsilon^j\mathcal{L}_{\pert{j}{\zeta}} \bar g_{\mu \nu} + \mathcal{O}\left(\epsilon^{j+1}\right).
\end{equation}
Of course, $\pert{i}{f}_{\ell\,\,00}$, $\pert{i}{f}_{\ell\,\,11}$,
$\pert{i}{f}_{\ell\,\,01}$, $\pert{i}{f}_{\ell\,\,+}$ so defined are not
gauge invariant in general; they change under gauge transformations
induced by ${}^{(j)}\zeta^{\mu}$ with $j < i$ (cf. Bruni \textit{et
  al.} \cite{BMMS}, Garat and Price \cite{GP}, eq. (25)). The RW gauge
corresponds to setting $\pert{i}{\zeta}_{\ell\,\,0} = \pert{i}{\zeta}_{\ell\,\,1} =
\pert{i}{\zeta}_{\ell\,\,2} = 0$ in (\ref{eq:h00}-\ref{eq:h-}). Other way
round, it can be easily seen that starting with any metric
perturbation of the form (\ref{eq:hPolar}), one can put
$\left(\pert{i}{h}_{22} - \pert{i}{h}_{33} / \sin^{2}\!\theta \right)$ to zero
with a suitable choice of $\pert{i}{\zeta}_{\ell\,\,2}$, then put $\pert{i}{h}_{12}$
to zero with a suitable choice of $\pert{i}{\zeta}_{\ell\,\,1}$, and finally
put $\pert{i}{h}_{02}$ to zero with a suitable choice of
$\pert{i}{\zeta}_{\ell\,\,0}$ rendering the metric perturbation in RW
gauge. At each nonlinear order ($i>1$) sources have the form
\begin{equation}
  \label{eq:S_order_i}
  \left( \pert{i}{S}_{\alpha\beta} \right) =
\left(\begin{array}{cccc}
    \pert{i}{S}_{00} & \pert{i}{S}_{01} & \pert{i}{S}_{02} & 0 \\
    \pert{i}{S}_{01} & \pert{i}{S}_{11} & \pert{i}{S}_{12} & 0 \\
    \pert{i}{S}_{02} & \pert{i}{S}_{12} & \pert{i}{S}_{22} & 0 \\
    0 & 0 & 0 & \pert{i}{S}_{33} \\
\end{array}\right) \, ,
\end{equation}
with the components expanded according to
(\ref{eq:polarT_ab}-\ref{eq:polarT-}). In the following we will use
extensively the fact that the sources $^{(i)}S_{\ell\,\,\mu\nu}$
fulfill three types of identities:
\begin{align}
\label{eq:zero0}
\pert{i}{\Nsc}_{\ell\,\,0} & := \frac{1}{2} \left( \frac{1}{\fsc} \partial_t \pert{i}{S}_{\ell\,\,00} + \fsc \partial_t \pert{i}{S}_{\ell\,\,11}\right) + \frac{1}{r^2}\partial_t \pert{i}{S}_{\ell\,\,+} - \fsc \partial_r \pert{i}{S}_{\ell\,\,01} - \frac{2 \fsc + r \, \fsc'}{r} \pert{i}{S}_{\ell\,\,01} + \frac{\ell(\ell+1)}{r^2} \pert{i}{S}_{\ell\,\,02} = 0\, ,
\\
\label{eq:zero1}
\pert{i}{\Nsc}_{\ell\,\,1} & := \frac{1}{2} \left(\frac{1}{\fsc} \partial_r \pert{i}{S}_{\ell\,\,00} + \fsc \partial_r \pert{i}{S}_{\ell\,\,11} \right) - \frac{1}{r^2}\partial_r \pert{i}{S}_{\ell\,\,+} - \frac{1}{\fsc} \partial_t \pert{i}{S}_{\ell\,\,01} + \frac{2 \fsc + r \, \fsc'}{r} \pert{i}{S}_{\ell\,\,11} - \frac{\ell(\ell+1)}{r^2} \pert{i}{S}_{\ell\,\,12} = 0\, ,
\\
\label{eq:zero2}
\pert{i}{\Nsc}_{\ell\,\,2} & := \frac{1}{2} \left( \frac{1}{\fsc} \, \pert{i}{S}_{\ell\,\,00} - \frac{1}{2} \fsc \, \pert{i}{S}_{\ell\,\,11}\right) - \frac{1}{\fsc} \partial_t \pert{i}{S}_{\ell\,\,02} + \fsc \partial_r \pert{i}{S}_{\ell\,\,12} + \frac{2\fsc + r \, \fsc'}{r} \pert{i}{S}_{\ell\,\,12} - \frac{(\ell-1)(\ell+2)}{r^2} \pert{i}{S}_{\ell\,\,-} = 0\, .
\end{align}
These are easily obtained by taking the background divergence of
(\ref{eq:pertEq}): $\pert{i}{\Nsc}_{\ell\,\,\nu} = 0$ (with $\nu=0,1,2$)
follows from (three) polar components of $\bar \nabla^{\mu}
\pert{i}{E}_{\mu\nu}=0$. We point out that the authors of \cite{dhs,ds} give
different identities: $\bar \nabla^{\mu} \pert{i}{S}_{\mu\nu}=0$ that are
apparently not correct in the case of polar type perturbations, as the
(three) polar components of $\bar \nabla^{\mu} \Delta_L \pert{i}{h}_{\mu
  \nu}$ are not identically zero.

The polar components of the system of perturbative Einstein equations
(\ref{eq:pertEq}) are also expanded according to
(\ref{eq:polarT_ab}-\ref{eq:polarT-}).  It is important to note that
the gauge degrees of freedom enter $\pert{i}{E}_{\ell\,\,\mu\nu}$ only
through the source terms $\pert{i}{S}_{\ell\,\,\mu\nu}$
($\pert{i}{S}_{\ell\,\,\mu\nu}$ depend on gauge functions
${}^{(j)}\zeta_{\ell\,\,0}$, ${}^{(j)}\zeta_{\ell\,\,1}$, ${}^{(j)}\zeta_{\ell\,\,2}$
with $j<i$, that is on the gauge choices made in previous steps). On
the other hand $\Delta_L \pert{i}{h}_{\ell\,\,\mu\nu}$ components are given
in terms of Regge-Wheeler gauge invariant variables
$\pert{i}{f}_{\ell\,\,00}$, $\pert{i}{f}_{\ell\,\,11}$, $\pert{i}{f}_{\ell\,\,01}$,
$\pert{i}{f}_{\ell\,\,+}$ only
\footnote{
We remind that here, and in the following, $\Delta_L
\pert{i}{h}_{\ell\,\,\mu\nu}$ refers to the expansion of the
\textit{tensor} $\Delta_L \pert{i}{h}_{\mu\nu}^{(i)}$ according to
(\ref{eq:polarT_ab}-\ref{eq:polarT-}) and not the action of the
$\Delta_L$ operator on $\pert{i}{h}_{\ell\,\,\mu\nu}(t,r)$~!}.
Thus we solve (\ref{eq:pertEq}) at order $i$ for Regge-Wheeler
variables $\pert{i}{f}_{\ell\,\,00}$, $\pert{i}{f}_{\ell\,\,11}$,
$\pert{i}{f}_{\ell\,\,01}$, $\pert{i}{f}_{\ell\,\,+}$ and then recover aAdS
gauge with a suitable gauge transformation (see below). It is also
known since the Regge and Wheeler seminal paper \cite{RW} that it is
convenient to search for the solutions of the system (\ref{eq:pertEq})
in terms of one scalar master variable. To achieve it we note that
\begin{equation}
\label{eq:algebraicE-}
\pert{i}{E}_{\ell\,\,-} = \frac{1}{4} \left( \frac{1}{\fsc} \, \pert{i}{f}_{\ell\,\,00}  - \fsc \, \pert{i}{f}_{\ell\,\,11} \right) - \pert{i}{S}_{\ell\,\,-}
\end{equation}
sets purely algebraic relation between $\pert{i}{f}_{\ell\,\,00}$ and $\pert{i}{f}_{\ell\,\,11}$, and the combination
\begin{align}
0 =& - \frac{1}{2} \left( \frac{1}{\fsc} \, \pert{i}{E}_{\ell\,\,00} - \fsc \, \pert{i}{E}_{\ell\,\,11} \right) +  \frac{1}{\fsc} \partial_t \pert{i}{E}_{\ell\,\,02}  - \fsc \partial_r \pert{i}{E}_{\ell\,\,12} - \frac{2 \fsc + r \, \fsc'}{r} \pert{i}{E}_{\ell\,\,12} + \frac{(\ell-1)(\ell+2)}{r^2} \pert{i}{E}_{\ell\,\,-}  
\nonumber\\
\label{eq:sourceIdentity}
& \equiv \frac{1}{2} \left( \frac{1}{\fsc} \, \pert{i}{S}_{\ell\,\,00} - \fsc \, \pert{i}{S}_{\ell\,\,11}^{(i)} \right) - \frac{1}{\fsc} \partial_t \pert{i}{S}_{\ell\,\,02}  + \fsc \partial_r \pert{i}{S}_{\ell\,\,12} + \frac{2\fsc + r \, \fsc' }{r} \pert{i}{S}_{\ell\,\,12} - \frac{(\ell-1)(\ell+2)}{r^2} \pert{i}{S}_{\ell\,\,-}
\end{align}
reduces to the identity (\ref{eq:zero2}) fulfilled by the
sources. Thus after eliminating $\pert{i}{f}_{\ell\,\,00}$ from
(\ref{eq:algebraicE-}) we are left with three RW gauge invariant
potentials $\pert{i}{f}_{\ell\,\,11}$, $\pert{i}{f}_{\ell\,\,01}$,
$\pert{i}{f}_{\ell\,\,+}$ to satisfy five linearly independent equations:
$\pert{i}{E}_{\ell\,\,00} = \pert{i}{E}_{\ell\,\,01} = \pert{i}{E}_{\ell\,\,+} =
\pert{i}{E}_{\ell\,\,02} = \pert{i}{E}_{\ell\,\,12} = 0$. All formulae
presented so far, hold for any spherically symmetric solution of vacuum
Einstein equations (i.e. Schwarzschild and Schwarzschild-(Anti-)de Sitter) in
Schwarzschild-like coordinates (\ref{eq:line_element}) with $\fsc=1
\pm (r^2/\lsc^2) - 2M/r$. Being interested in perturbations of AdS
space, from now on we restrict to maximally symmetric solutions
($M=0$), as in this case apart form the general identity $r^2 \fsc''= 2
\fsc - 2$ we have also an identity $r \fsc'=2\fsc - 2$ (valid only in
$M=0$ case) and it slightly simplifies discussion that follows. We
introduce the master scalar variable for polar perturbations at linear
order in such a way, to make the equations $\pert{1}{E}_{\ell\,\,02} =
\pert{1}{E}_{\ell\,\,12} = 0$ identically satisfied. Namely, writing down
the potentials $\pert{1}{f}_{\ell\,\,11}$, $\pert{1}{f}_{\ell\,\,01}$,
$\pert{1}{f}_{\ell\,\,+}$ at linear order in terms of linear combination
of the derivatives (up to second order) of \textit{one} master scalar
variable for polar perturbations $\pert{1}{\Phi}^{\Psc}_{\ell}$, and then
plugging this combination into $\pert{1}{E}_{\ell\,\,02} =
\pert{1}{E}_{\ell\,\,12} = 0$ yields a unique such combination (up to a
multiplicative factor):
\begin{align}
\label{eq:Muk00}
\pert{1}{f}_{\ell\,\,00} = & \frac{r}{2} \left( \partial_{tt} \pert{1}{\Phi}^{\Psc}_{\ell} + \fsc^2 \partial_{rr} \pert{1}{\Phi}^{\Psc}_{\ell} \right) + \fsc^2 \partial_{r} \pert{1}{\Phi}^{\Psc}_{\ell} = \fsc^2 \, \pert{1}{f}_{\ell\,\,11} \, ,
\\
\label{eq:Muk01}
\pert{1}{f}_{\ell\,\,01} = & r \partial_{rt} \pert{1}{\Phi}^{\Psc}_{\ell} + \fsc^{-1} \partial_{t} \pert{1}{\Phi}^{\Psc}_{\ell} \, ,
\\
\label{eq:MukY}
\pert{1}{f}_{\ell\,\,+} = & -\frac{r}{2} \left( \fsc^{-1} \partial_{tt} \pert{1}{\Phi}^{\Psc}_{\ell} - \fsc \partial_{rr} \pert{1}{\Phi}^{\Psc}_{\ell} \right) + \left(2 \fsc - 1\right) \partial_{r} \pert{1}{\Phi}^{\Psc}_{\ell} \, .
\end{align}
This is an easy way to recover old results of Mukohyama
\cite{Mukohyama}, eq.~(44) and Kodama and Ishibashi \cite{KI},
eq.~(3.14). At linear order the three remaining equations
$\pert{1}{E}_{\ell\,\,00} = \pert{1}{E}_{\ell\,\,01} = \pert{1}{E}_{\ell\,\,+} = 0$
are satisfied if and only if $\pert{1}{\Phi}^{\Psc}_{\ell}$ solves
homogeneous radial wave equation 
\footnote{Strictly speaking 
\[
\tilde \Box_{\ell} \Phi(t,r) = r \left(-\bar \Box + V_{\ell} \right) \frac{\Phi(t,r)}{r} \,,
\]
where $\bar \Box$ is the wave operator for the metric (\ref{eq:line_element}) and the potential $V_{\ell}$ reads
\[
V_{\ell} = \frac{\ell(\ell+1)}{r^2} - \frac{\fsc'}{r} \,.
\]
}:
\begin{equation}
\label{eq:H_scalar_wave}
\tilde \Box_{\ell} \pert{1}{\Phi}^{\Psc}_{\ell} := \frac{1}{\fsc} \partial_{tt} \pert{1}{\Phi}^{\Psc}_{\ell} - \fsc \, \partial_{rr} \pert{1}{\Phi}^{\Psc}_{\ell} - \fsc' \, \partial_{r} \pert{1}{\Phi}^{\Psc}_{\ell} + \frac{\ell(\ell+1)}{r^2} \, \pert{1}{\Phi}^{\Psc}_{\ell} = 0 \,,
\end{equation}
namely under the substitution (\ref{eq:Muk00}-\ref{eq:MukY})
\begin{align}
\label{eq:E00Box}
\pert{1}{E}_{\ell\,\,00} =& \frac{\fsc^2}{4} \left( \frac{1}{\fsc^2} \partial_{tt} + r \partial_{rr} + \partial_{r} + \frac{6}{r} \right) \tilde \Box_{\ell} \pert{1}{\Phi}^{\Psc}_{\ell} \, ,
\\
\label{eq:E01Box}
\pert{1}{E}_{\ell\,\,01} =& \left( \frac{r}{2} \partial_{r} + 1 + \frac{1}{2 \fsc} \right) \partial_t \tilde \Box_{\ell} \pert{1}{\Phi}^{\Psc}_{\ell} \, ,
\\
\label{eq:EPlusBox}
\pert{1}{E}_{\ell\,\,+} =& \left( - \frac{r^3}{4 \fsc} \partial_{tt} + \frac{r^3 \fsc}{4}\partial_{rr} + r^2 \left(2 \fsc - \frac{1}{2}\right) \partial_{r} + r \left(\frac{5}{2} \fsc  - 1\right) \right) \tilde \Box_{\ell} \pert{1}{\Phi}^{\Psc}_{\ell} \, .
\end{align}
To proceed to higher orders it is convenient to invert
(\ref{eq:Muk00}-\ref{eq:MukY}) for the master scalar variable
$\pert{1}{\Phi}^{\Psc}_{\ell}$. We take the following linear
combination of RW gauge invariant potentials $\pert{i}{f}_{\ell\,\,11}$,
$\pert{i}{f}_{\ell\,\,01}$, $\pert{i}{f}_{\ell\,\,+}$ and their first
derivatives:
\begin{equation}
\label{eq:Phi_def}
\pert{i}{\Phi}^{\Psc}_{\ell} = \frac{2 r}{\ell(\ell+1)} \left( \pert{i}{f}_{\ell\,\,+}  + 2 \fsc \, \frac{\fsc \, \pert{i}{f}_{\ell\,\,11} - r \partial_r \pert{i}{f}_{\ell\,\,+}}{(\ell-1)(\ell+2)}\right) \,.
\end{equation}
Indeed at linear order, under substitution
(\ref{eq:Muk00}-\ref{eq:MukY}) with (\ref{eq:H_scalar_wave}) fulfilled
(\ref{eq:Phi_def}) becomes an identity. If translated to the
asymptotically flat Schwarzschild case this definition would
corresponds to Brizuela et al. \cite{BMGT}, eq.~(11) , that dates back
to the choice made by Moncrief \cite{Moncrief}.  We take
(\ref{eq:Phi_def}) as a \textit{definition} of master scalar variable
also at higher (nonlinear) orders. This definition determines the
source for the inhomogeneous wave equation for master scalar variable
at higher orders, namely substituting (\ref{eq:Phi_def}) on the RHS of
(\ref{eq:H_scalar_wave}) we find that
\begin{align}
& \tilde \Box_{\ell} \pert{i}{\Phi}^{\Psc}_{\ell} = \frac {4 r^2} {(\ell-1)\ell(\ell+1)(\ell+2)} 
\nonumber\\
\times & \left( (\fsc / r) \left( \fsc \Delta_L \pert{i}{h}_{\ell\,\,11} - \frac{1}{\fsc} \Delta_L \pert{i}{h}_{\ell\,\,00} \right) + \frac{(\ell-1)(\ell+2) - 2 \left(3\fsc - 2\right)}{r^3} \Delta_L \pert{i}{h}_{\ell\,\,+}  - 2 \fsc \partial_r \left( \Delta_L \pert{i}{h}_{\ell\,\,+} / r^2 \right) \right.
\nonumber\\
& \hskip 1mm \left. - \frac{2 \ell (\ell+1)}{r^2} \fsc \Delta_L \pert{i}{h}_{\ell\,\,12} + \frac{(\ell-1)\ell(\ell+1)(\ell+2)}{r^3} \Delta_L \pert{i}{h}_{\ell\,\,-} \right) \, .
\end{align}  
Thus at at higher orders ($i \geq 2$) the definition (\ref{eq:Phi_def}) leads to  
\begin{equation}
\label{eq:NH_scalar_wave}
\pert{i}{\Bsc}^{\Psc}_{\ell} :=\tilde \Box_{\ell} \pert{i}{\Phi}^{\Psc}_{\ell} - \pert{i}{\tilde S}^{\Psc}_{\ell} = 0\, , 
\end{equation}
with
\begin{align}
\label{eq:source_scalar_wave}
\pert{i}{\tilde S}^{\Psc}_{\ell} = & \frac {4 r^2} {(\ell-1)\ell(\ell+1)(\ell+2)} 
\nonumber\\
\times & \left( (\fsc / r) \left( \fsc \, \pert{i}{S}_{\ell\,\,11} - \frac{1}{\fsc} \, \pert{i}{S}_{\ell\,\,00} \right) + \frac{(\ell-1)(\ell+2) - 2 \left( 3 \fsc - 2 \right)}{r^3} \pert{i}{S}_{\ell\,\,+} - 2 \fsc \partial_r \left( \left. \pert{i}{S}_{\ell\,\,+} \right/ r^2 \right) \right.
\nonumber\\
& \hskip 1mm \left. - \frac{2 \ell (\ell+1)}{r^2} \fsc \, \pert{i}{S}_{\ell\,\,12} + \frac{(\ell-1)\ell(\ell+1)(\ell+2)}{r^3} \pert{i}{S}_{\ell\,\,-} \right)
\end{align}
for $\ell \geq 2$ (here the $\ell=0$ and $\ell=1$ cases have to be
treated separately). The linear rules (\ref{eq:Muk00}-\ref{eq:MukY})
are generalized to
\begin{align}
\label{eq:NHrules00tmp}
\pert{i}{f}_{\ell\,\,00} = & \fsc^2 \left( \frac{r}{2} \left( \frac{1}{\fsc^2} \partial_{tt} \pert{i}{\Phi}^{\Psc}_{\ell} + \partial_{rr} \pert{i}{\Phi}^{\Psc}_{\ell} \right) + \partial_{r} \pert{i}{\Phi}^{\Psc}_{\ell} + \pert{i}{\alpha}_{\ell}(t,r) \right) + 4 \fsc \, \pert{i}{S}_{\ell\,\,-} \, ,
\\
\label{eq:NHrules11tmp}
\pert{i}{f}_{\ell\,\,11} = & \hskip 7mm \frac{r}{2} \left( \frac{1}{\fsc^2} \partial_{tt} \pert{i}{\Phi}^{\Psc}_{\ell} + \partial_{rr} \pert{i}{\Phi}^{\Psc}_{\ell} \right) + \partial_{r} \pert{i}{\Phi}^{\Psc}_{\ell} + \pert{i}{\alpha}_{\ell}(t,r) \, ,
\\
\label{eq:NHrules01tmp}
\pert{i}{f}_{\ell\,\,01} = & r \partial_{rt} \pert{i}{\Phi}^{\Psc}_{\ell} + \fsc^{-1} \partial_{t} \pert{i}{\Phi}^{\Psc}_{\ell} + \pert{i}{\beta}_{\ell}(t,r) \, ,
\\
\label{eq:NHrulesYtmp}
\pert{i}{f}_{\ell\,\,+} = & -\frac{r}{2} \left( \frac{1}{\fsc} \partial_{tt} \pert{i}{\Phi}^{\Psc}_{\ell} - \fsc \partial_{rr} \pert{i}{\Phi}^{\Psc}_{\ell} \right) + \left(2 \fsc - 1\right) \partial_{r} \pert{i}{\Phi}^{\Psc}_{\ell} + \pert{i}{\gamma}_{\ell}(t,r) \, ,
\end{align}
with the set of three functions $\pert{i}{\alpha}_{\ell}$,
$\pert{i}{\beta}_{\ell}$ and $\pert{i}{\gamma}_{\ell}$ for each $\ell$, that
still have to be specified. These functions are set in such a way that
the equations $\pert{i}{E}_{\ell\,\,02} = \pert{i}{E}_{\ell\,\,12} = 0$ are
identically satisfied, while the equations $\pert{i}{E}_{\ell\,\,00} =
\pert{i}{E}_{\ell\,\,01} = \pert{i}{E}_{\ell\,\,+} =0$ are satisfied if and
only if $\pert{i}{\Phi}^{\Psc}_{\ell}$ solves the inhomogeneous wave
equation (\ref{eq:NH_scalar_wave}), i.e.
\begin{align}
\label{eq:E00BoxNH}
\pert{i}{E}_{\ell\,\,00} =& \frac{\fsc^2}{4} \left( \frac{r}{\fsc^2} \partial_{tt} + r \partial_{rr} + \partial_{r} + \frac{6}{r} \right) \pert{i}{\Bsc}^{\Psc}_{\ell} \, ,
\\
\label{eq:E01BoxNH}
\pert{i}{E}_{\ell\,\,01} =& \left( \frac{r}{2} \partial_{r} + 1 + \frac{1}{2 \fsc} \right) \partial_t \pert{i}{\Bsc}^{\Psc}_{\ell} \, ,
\\
\label{eq:EPlusBoxNH}
\pert{i}{E}_{\ell\,\,+} =& \left( - \frac{r^3}{4 \fsc} \partial_{tt} + \frac{r^3 \fsc}{4}\partial_{rr} + r^2 \left(2 \fsc - \frac{1}{2}\right) \partial_{r} + r \left(\frac{5}{2} \fsc - 1\right) \right) \pert{i}{\Bsc}^{\Psc}_{\ell} \, ,
\end{align}
cf. the linear case (\ref{eq:E00Box}-\ref{eq:EPlusBox}). It is worth
to note that due to the identities (\ref{eq:zero0}-\ref{eq:zero2}) the
solutions for $\pert{i}{\alpha}_{\ell}$, $\pert{i}{\beta}_{\ell}$ and
$\pert{i}{\gamma}_{\ell}$ can be found in purely algebraic way, without
need to solve any (partial) differential equations. The results read:
\begin{align}
\label{eq:alpha}
\pert{i}{\alpha}_{\ell} &= -\frac {2 r^2 \left( \frac{\displaystyle 1}{\displaystyle \fsc} \, \pert{i}{S}_{\ell\,\,00} - \fsc \, \pert{i}{S}_{\ell\,\,11} \right) - \frac{\displaystyle r^2}{\displaystyle \fsc} (\ell-1)(\ell+2) \left( \frac{1}{\fsc} \, \pert{i}{S}_{\ell\,\,00} + \fsc \, \pert{i}{S}_{\ell\,\,11} \right) + 4 \partial_r \left( r \, \pert{i}{S}_{\ell\,\,+} \right) + \frac{\displaystyle 4 r^3}{\displaystyle \fsc} \partial_t \pert{i}{S}_{\ell\,\,01}} {(\ell-1)\ell(\ell+1)(\ell+2)}  
\nonumber\\
&- \frac{2}{\fsc} S_{\ell\,\,-} \, ,
\\
\label{eq:beta}
\pert{i}{\beta}_{\ell} &= \frac {r \ell (\ell+1) \pert{i}{S}_{\ell\,\,02} - r^2 \left( 4 \fsc -1 - \ell(\ell+1)/2 \right) \pert{i}{S}_{\ell\,\,01} - r^3 \fsc \partial_r \pert{i}{S}_{\ell\,\,01}} {(\ell-1)\ell(\ell+1)(\ell+2)/4} \, ,
\\
\label{eq:gamma}
\pert{i}{\gamma}_{\ell} &= -\frac {r^2 \, \pert{i}{S}_{\ell\,\,00} + r \fsc \ell (\ell+1) \pert{i}{S}_{\ell\,\,12} + \left( 2 \fsc - 1 - \ell(\ell+1)/2 \right) \pert{i}{S}_{\ell\,\,+} + r \fsc \partial_r \pert{i}{S}_{\ell\,\,+}} {(\ell-1)\ell(\ell+1)(\ell+2)/4} + 2 \, \pert{i}{S}_{\ell\,\,-} \, .
\end{align}
We stress that the Regge-Wheeler gauge (i.e. setting
$\pert{i}{\zeta}_{\ell\,\,0} = \pert{i}{\zeta}_{\ell\,\,1} = \pert{i}{\zeta}_{\ell\,\,2} = 0$
in (\ref{eq:h00}-\ref{eq:h-})) is not aAdS (nor
asymptotically flat in the corresponding $\Lambda=0$ case). The
asymptotic form of vacuum metric perturbations corresponding to an aAdS
space was found in \cite{HenneauxTeitelboim}. Here we follow the
discussion in Sec.~II.C of \cite{BantilanPretoriusGubser}. We ensure
that $O(2,3)$ is an asymptotic symmetry of $g_{\mu\nu} = \bar
g_{\mu\nu} + \delta g_{\mu\nu}$ by requiring the Killing equation to
be satisfied in asymptotic sense
\begin{equation}
\label{eq:aAdS1}
\mathcal{L}_{\xi} g_{\mu\nu} = \mathcal{O} \left( \delta g_{\mu\nu}
\right)
\end{equation}
for all Killing vectors of AdS space. Thus we are looking for an
asymptotic form of metric perturbations $\delta g_{\mu\nu} \sim
1/r^{\gamma_{\mu\nu}}$, such that this asymptotic form is preserved by
coordinate transformations generated by any AdS Killing vector
$\xi$. This holds for
\begin{equation}
\label{eq:aAdS2}
\gamma_{rr}=5, \mbox{ } \gamma_{r\mu}=4 \mbox{ and } \gamma_{\mu\nu}=1
\mbox{ for } \mu,\nu \neq r
\end{equation}
(in $3+1$ dimensions)
\cite{HenneauxTeitelboim,BantilanPretoriusGubser}. The asymptotic
behavior of solutions of homogeneous wave equation $\tilde \Box_{\ell} \Phi_{\ell} =0$
(\ref{eq:H_scalar_wave}) reads
\begin{equation}
\label{eq:Box_asymptotics}
\Phi_{\ell} \sim a_{\ell} + \frac{b_{\ell}}{r} +
  \mathcal{O}\left(\frac{1}{r^2}\right) \,,
\end{equation}
thus already at linear order RW variables reconstructed from
(\ref{eq:Muk00}-\ref{eq:MukY}) do not have AdS asymptotics and the RW
gauge is not aAdS. This is not a problem as long as a suitable gauge
transformation (i.e. suitable functions $\pert{i}{\zeta}_{\ell\,\,0},
\pert{i}{\zeta}_{\ell\,\,1}, \pert{i}{\zeta}_{\ell\,\,2}$) to render metric
perturbation $\pert{i}{h}_{\mu \nu}$ in aAdS form can be found. The
necessary condition for such gauge transformation to exist for polar type
perturbation is $b_{\ell} \equiv 0$ in (\ref{eq:Box_asymptotics}),
  that is the correct boundary condition at infinity for master scalar
  variable for polar perturbations $\pert{i}{\Phi}^{\Psc}_{\ell}$ reads
\begin{equation}
\label{eq:polarBC}
\pert{i}{\Phi}^{\Psc}_{\ell} = \pert{i}{a}_{\ell} + \mathcal{O}\left(\frac{1}{r^2}\right)\, 
\end{equation}
(this correct form of asymptotics was stressed for the first time in
\cite{dhs}). After separating time dependence in
(\ref{eq:NH_scalar_wave}) in the form $\pert{i}{\Phi}^{\Psc}_{\ell}(t,r)
= \sum_k \pert{i}{\Phi}^{\Psc}_{\ell\,\,k}(r) \cos\left(\omega_k t +
\phi_k\right)$ (with some suitable set of frequencies $\omega_k$) we
end up with ordinary second order differential equations for
$\pert{i}{\Phi}^{\Psc}_{\ell\,\,k}(r)$ coefficients, thus to obtain a
unique solution we need two boundary conditions; these are smoothness
in the center and (\ref{eq:polarBC}). To get the asymptotics of the
sources $\pert{i}{S}^{\Psc}_{\ell}$ in (\ref{eq:NH_scalar_wave})
compatible with the asymptotics of the~$\tilde \Box_{\ell}$~operator
(\ref{eq:Box_asymptotics}) (to allow for an easy control over aAdS
asymptotics step after step in perturbation expansion) it is necessary
to keep metric perturbations in aAdS form at each order. Once the
boundary condition (\ref{eq:polarBC}) is satisfied the aAdS gauge can
be recovered with the polar gauge vector induced by
(cf. (\ref{eq:polarV_a},\ref{eq:polarV_2}))
\begin{align}
\label{eq:gaugeM1}
\pert{i}{\zeta}_{\ell\,\,1} &= \frac{1}{4} \left( \lsc^2 \fsc^{-1} \partial_{tt} \pert{i}{\Phi}^{\Psc}_{\ell} - 4 r \partial_{r} \pert{i}{\Phi}^{\Psc}_{\ell} - r^2 \partial_{rr} \pert{i}{\Phi}^{\Psc}_{\ell} \right) \, ,
\\
\pert{i}{\zeta}_{\ell\,\,0} &= -r \partial_t \pert{i}{\zeta}_{\ell\,\,1} + \frac{r}{3 \fsc} \partial_t \left( \lsc^2 \partial_{tt} \pert{i}{\Phi}^{\Psc}_{\ell} + \pert{i}{\Phi}^{\Psc}_{\ell} \right) \, ,
\\
\label{eq:gaugeM2}
\pert{i}{\zeta}_{\ell\,\,2} &= \frac{r}{3} \, \pert{i}{\zeta}_{\ell\,\,1} \, .
\end{align}
In fact at first and second order we found also somewhat simpler rules:
\begin{align}
\label{eq:gaugeM1_1st}
{}^{(1)}\zeta_{\ell\,\,1} &= \frac{1}{4} \left( \lsc^2 \fsc^{-1} \partial_{tt} \pert{1}{\Phi}^{\Psc}_{\ell} - r \partial_{r} \pert{1}{\Phi}^{\Psc}_{\ell} \right) \, ,
\\
{}^{(1)}\zeta_{\ell\,\,0} &= -r \partial_t {}^{(1)}\zeta_{\ell\,\,1} + \frac{r}{3 \fsc} \partial_t \left( \lsc^2 \partial_{tt} \pert{1}{\Phi}^{\Psc}_{\ell} + \pert{1}{\Phi}^{\Psc}_{\ell} \right) \, ,
\\
\label{eq:gaugeM2_1st}
{}^{(1)}\zeta_{\ell\,\,2} &= \frac{r}{3} \, {}^{(1)}\zeta_{\ell\,\,1} \, ,
\end{align} 
and
\begin{align}
\label{eq:gaugeM02nd}
{}^{(2)}\zeta_{\ell\,\,0} &= 0 \, ,
\\
{}^{(2)}\zeta_{\ell\,\,1} &= \frac{1}{2} \left( {}^{(2)}\Phi^{\Psc}_{\ell} - \frac{1}{2} \partial_{rr}\left( r^2 \, {}^{(2)}\Phi^{\Psc}_{\ell}\right) \right) \, ,
\\
\label{eq:gaugeM12nd}
{}^{(2)}\zeta_{\ell\,\,2} &= \frac{r}{3} \, {}^{(2)}\zeta_{\ell\,\,1} \, .
\end{align} 

To summarize: to satisfy the set of perturbative Einstein equations
(\ref{eq:pertEq}) for polar type perturbations at axial symmetry, at
any nonlinear order, for angular momenta $\ell \geq 2$
(cf. (\ref{eq:polarT_ab}-\ref{eq:polarT-})), it is enough to solve just one
inhomogeneous wave equation for scalar master variable for polar
perturbations (\ref{eq:NH_scalar_wave}) with the source term given in
(\ref{eq:source_scalar_wave}), and then reconstruct the Regge-Wheeler
gauge invariant potentials $\pert{i}{f}_{\ell\,\,00}$,
$\pert{i}{f}_{\ell\,\,11}$, $\pert{i}{f}_{\ell\,\,01}$ and
$\pert{i}{f}_{\ell\,\,+}$ according to
(\ref{eq:NHrules00tmp}-\ref{eq:NHrulesYtmp}) together with
(\ref{eq:alpha}-\ref{eq:gamma}). Then with the gauge transformation
(\ref{eq:gaugeM1}-\ref{eq:gaugeM2}) the resulting perturbations
$\pert{i}{h}_{\ell\,\,\mu\nu}$ (cf. (\ref{eq:h00}-\ref{eq:h-})) can be put
in asymptotically AdS form.

As was stressed above the special cases $\ell=0$ and $\ell=1$ need a special treatment.

\subsection{The $\ell=0$ case for polar perturbations.}
In the $\ell=0$ case the only nontrivial equations are $0 = \pert{i}{E}_{0\,\,+} = \pert{i}{E}_{0\,\,01} = \pert{i}{E}_{0\,\,1 \pm 0} := \left(\fsc \pert{i}{E}_{0\,\,11} \pm (1 / \fsc) \pert{i}{E}_{0\,\,00}\right)$. In fact 
\begin{align}
\pert{i}{E}_{0\,\,1 - 0} -\frac{1}{r} \partial_r \pert{i}{E}_{0\,\,+} + \frac{r}{2} \partial_r \pert{i}{E}_{0\,\,1 + 0} - \frac{r}{\fsc} \partial_t \pert{i}{E}_{0\,\,01} +  \frac{2 \fsc - 1}{\fsc} \pert{i}{E}_{0\,\,1 + 0} & \equiv 0, 
\\
\partial_t \left( \pert{i}{E}_{0\,\,+} + \frac{r^2}{2} \pert{i}{E}_{0\,\,1 + 0} \right) - 2r \left(2 \fsc - 1\right) \pert{i}{E}_{0\,\,01} - r^2 \fsc \partial_r \pert{i}{E}_{0\,\,01} & \equiv 0
\end{align}
(due to identities (\ref{eq:zero0},\ref{eq:zero1})), and $\pert{i}{f}_{0\,\,01}$, $\pert{i}{f}_{0\,\,+}$ become gauge degrees of freedom (i.e. they can be put to zero with a suitable choice of $\pert{i}{\zeta}_{0\,\,0}$ and $\pert{i}{\zeta}_{0\,\,1}$). Thus we are left with two equations $0 = \pert{i}{E}_{0\,\,0,1} = \pert{i}{E}_{0\,\,1 + 0}$ for two unknown functions $\pert{i}{f}_{0\,\,00}$ and $\pert{i}{f}_{0\,\,11}$, where $\pert{i}{f}_{0\,\,01}$ and $\pert{i}{f}_{0\,\,+}$ can be freely specified and we put them to zero. This system can be easily integrated to yield 
\begin{align}
\label{eq:f11ell0}
\pert{i}{f}_{0\,\,11} &= \pert{i}{f}^{\Psc \,\, res}_{0} + r \fsc^{-1} \int^t \pert{i}{S}_{0\,\,01} \, dt'
\\
\label{eq:f00ell0}
\pert{i}{f}_{0\,\,00} &= \fsc^2 \, \pert{i}{f}_{0\,\,11} - \fsc \int_{\infty}^r r' \left( \pert{i}{S}_{0\,\,11} + \frac{1}{\fsc^2} \, \pert{i}{S}_{0\,\,00} \right)\, dr' \,,
\end{align}  
with the residual degree of freedom  $\pert{i}{f}^{\Psc \,\, res}_{0} \equiv \pert{i}{f}^{\Psc \,\, res}_{0}(r)$ that is not set by the equations $0 = E^{(i)}_{0\,\,0,1} = E^{(i)}_{0\,\,1 + 0}$. It is however uniquely determined as the solution of the first order ordinary differential equation set by (the time independent part of) the equation $\pert{i}{E}_{0\,\,+}=0$. The solution reads
\begin{equation}
\pert{i}{f}^{\Psc \,\, res}_{0} = \frac{1}{r \, \fsc^2} \int_0^r \left[ \frac{r'^2}{2} \left( \frac{1}{\fsc} \, \pert{i}{\bar S}_{0\,\,00} + \fsc \, \pert{i}{\bar S}_{0\,\,11}\right) +  \pert{i}{\bar S}_{0\,\,+} \right] \, dr' \, , 
\end{equation}
where $\pert{i}{\bar S}_{\ell\,\,\mu\nu}$ are time independent parts of $\pert{i}{S}_{\ell\,\,\mu\nu}$ and $\pert{i}{f}^{\Psc \,\, res}_{0}$ is thus by definition time independent. At linear order the $\ell=0$ part of the perturbation reduces to a pure gauge, as expected (as ${}^{(1)}S_{\mu\nu} \equiv 0$). Moreover for $i>1$, $\pert{i}{f}_{0\,\,00}$ and $\pert{i}{f}_{0\,\,11}$ satisfy aAdS conditions (\ref{eq:aAdS1},\ref{eq:aAdS2}), thus no further gauge transformation is needed.

\subsection{The $\ell=1$ case for polar perturbations.} 
In the $\ell=1$ case we proceed similarly to the $\ell=0$ case. There are six nontrivial Einstein equation: $0 = \pert{i}{E}_{1\,\,+} = \pert{i}{E}_{1\,\,01} = \pert{i}{E}_{1\,\,02} = \pert{i}{E}_{1\,\,12} = \pert{i}{E}_{1\,\,1 \pm 0} := \left(\fsc \pert{i}{E}_{1\,\,11} \pm (1 / \fsc)\pert{i}{E}_{1\,\,00}\right)$. In fact
\begin{eqnarray}
\frac{1}{2} \, \pert{i}{E}_{1\,\,1-0} +  \frac{1}{\fsc} \partial_t \pert{i}{E}_{1\,\,02}  - \fsc \partial_r \pert{i}{E}_{1\,\,12} - \frac{2}{r} \left(2 \fsc - 1\right) \pert{i}{E}_{1\,\,12} & \equiv 0 \, , 
\\
\partial_t \left( \pert{i}{E}_{1\,\,+} + \frac{r^2}{2} \pert{i}{E}_{1\,\,1 + 0} \right) - 2r \left(2 \fsc - 1\right) \pert{i}{E}_{1\,\,01} - r^2 \fsc \partial_r \pert{i}{E}_{1\,\,01} + 2 \, \pert{i}{E}_{1\,\,02} & \equiv 0 \, ,
\\
-\frac{1}{r} \partial_r \pert{i}{E}_{1\,\,+} + \frac{r}{2} \partial_r \pert{i}{E}_{1\,\,1 + 0} + \frac{2 \fsc - 1}{\fsc} \pert{i}{E}_{1\,\,1 + 0} - \frac{r}{\fsc} \partial_t \pert{i}{E}_{1\,\,01} - \frac{2}{r} \pert{i}{E}_{1\,\,12} + \pert{i}{E}_{1\,\,1-0} & \equiv 0
\end{eqnarray} 
due to identities (\ref{eq:zero0}-\ref{eq:zero2}), and $\pert{i}{f}_{1\,\,+}$ becomes gauge degree of freedom (i.e. it can be put to zero with a suitable choice of $\pert{i}{\zeta}_{1\,\,1}$ in function of  $\pert{i}{\zeta}_{1\,\,2}$). We set $\pert{i}{f}_{1\,\,+}$ to zero and integrate in sequence the equations $0 = r \, \pert{i}{E}_{1\,\,01} + 2 \,  \pert{i}{E}_{1\,\,02} = \pert{i}{E}_{1\,\,+} + (r^2/2) \, \pert{i}{E}_{1\,\,1+0} = \pert{i}{E}_{1\,\,12}$ to get
\begin{align}
\label{eq:f01ell1}
\pert{i}{f}_{1\,\,01} &= \frac{1}{r \, \fsc^{1/2}} \int_0^r \frac{r'}{\fsc^{1/2}} \left( r'  \, \pert{i}{S}_{1\,\,01} + 2 \, \pert{i}{S}_{1\,\,02} \right) \, dr' \, ,
\\
\label{eq:f11ell1}
\pert{i}{f}_{1\,\,11} &= \frac{1}{r^2 \, \fsc^{3/2}} \int_0^r \frac{r'}{\fsc^{1/2}} \left[ \frac{r'^2}{2} \left( \frac{1}{\fsc} \, \pert{i}{S}_{1\,\,00} + \fsc \, S_{1\,\,11} \right) + \pert{i}{S}_{1\,\,+} \right] \, dr' \, , 
\\
\label{eq:f00ell1}
\pert{i}{f}_{1\,\,00} &= r \fsc^{1/2} \int_{\infty}^r \frac{\fsc^{1/2}}{r'} \left( 2 \, \pert{i}{S}_{1\,\,12} + \frac{1}{\fsc} \partial_t \pert{i}{f}_{1\,\,01} - \frac{2 \fsc - 1}{r'} \pert{i}{f}_{1\,\,11} \right)\, dr' \, 
\end{align}  
respectively. Then equations $0 = \pert{i}{E}_{1\,\,01} = \pert{i}{E}_{1\,\,+}$ are also satisfied due to identities (\ref{eq:zero0}-\ref{eq:zero2}). Alternatively one can integrate the $\pert{i}{E}_{1\,\,01} = 0$ equation (instead of the $\pert{i}{E}_{1\,\,+} + (r^2/2) \, \pert{i}{E}_{1\,\,1+0} = 0$ equation) up to a residual $\pert{i}{f}^{\Psc \,\, res}_{1} \equiv \pert{i}{f}^{\Psc \,\, res}_{1}(r)$:
\begin{equation}
\label{eq:f11ell1Altenative}
\pert{i}{f}_{1\,\,11}^{(i)} = \pert{i}{f}^{\Psc \,\, res}_{1} + \frac{1}{\fsc} \int^t \left(r \, \pert{i}{S}_{1\,\,01} - \frac{1}{r} \, \pert{i}{f}_{1\,\,01} \right)\, dt' \, ,
\end{equation}
and then set $\pert{i}{f}^{\Psc \,\, res}_{1}$ by (the time independent part of) the equation $\pert{i}{E}^{(i)}_{1\,\,+} + (r^2 / 2) \, \pert{i}{E}_{1\,\,1 + 0}=0$:
\begin{equation}
\pert{i}{f}^{\Psc \,\, res}_{1} = \frac{1}{r^2 \, \fsc^{3/2}} \int_0^r \frac{r'}{\fsc^{1/2}} \left[ \frac{r'^2}{2} \left( \frac{1}{\fsc} \, \pert{i}{\bar S}_{1\,\,00} + \fsc \, \pert{i}{\bar S}_{1\,\,11}\right) +  \pert{i}{\bar S}_{1\,\,+} \right] \, dr' \, .
\end{equation}
Then the equations $0=\pert{i}{E}_{1\,\,+}=\pert{i}{E}_{1\,\,1 + 0}$ are also satisfied due to identities (\ref{eq:zero0}-\ref{eq:zero2}). This is a bit more convenient in some practical applications, as for the sources generated by time periodic solutions taking integrals in time is straightforward. At linear order the $\ell=1$ part of the perturbation reduces to a pure gauge, as expected (as ${}^{(1)}S_{\mu\nu} \equiv 0$). Moreover for $i>1$, $\pert{i}{f}_{1\,\,01}$, $\pert{i}{f}_{1\,\,00}$ and $\pert{i}{f}_{1\,\,11}$ satisfy aAdS condition (\ref{eq:aAdS1},\ref{eq:aAdS2}), thus no further gauge transformation is needed.

\section{Axial perturbations at axial symmetry}

The axial type perturbations can be treated along the same lines as
polar type perturbations discussed in the previous section, but are in
fact much easier to deal with at a technical level. We describe axial
type perturbations in this section for the sake of completeness. In
axial symmetry for axial type perturbations we have
\begin{equation}
\label{hAxial_in_RW}
\left( \pert{i}{h}_{\alpha\beta} \right)= 
\left(\begin{array}{cccc}
    0 & 0 & 0 & \pert{i}{h}_{03} \\
    0 & 0 & 0 & \pert{i}{h}_{13} \\
    0 & 0 & 0 & \pert{i}{h}_{23} \\
    \pert{i}{h}_{03} & \pert{i}{h}_{13} & \pert{i}{h}_{23} & 0 
\end{array}\right)\, ,
\end{equation}
with (cf. (\ref{eq:axialT_a3},\ref{eq:axialT_23}))
\begin{align}
\label{eq:h03}
\pert{i}{h}_{\ell\,\,03} = & \pert{i}{f}_{\ell\,\,03} + \partial_t \pert{i}{\eta}_{\ell} \, ,
\\
\pert{i}{h}_{\ell\,\,13} = & \pert{i}{f}_{\ell\,\,13} + \partial_r \pert{i}{\eta}_{\ell} - 2 \, \pert{i}{\eta}_{\ell} / r \, ,
\\
\label{eq:h23}
\pert{i}{h}_{\ell\,\,23} = & \pert{i}{\eta}_{\ell} \, , 
\end{align}
where an axial component $\pert{i}{\eta}_{\ell}(t,r)$ defines the $i$-th order axial gauge vector $\pert{i}{\eta}_{\mu}$ (cf. (\ref{eq:axialV_3})) and $\pert{i}{f}_{\ell\,\,03}(t,r)$, $\pert{i}{f}_{\ell\,\,13}(t,r)$ are Regge-Wheeler variables \cite{RW, Nollert} being gauge invariant with respect to gauge transformations induced by ${}^{(j)}\eta_{\mu}$ with $j \geq i$, cf. (\ref{eq:gaugeTransform}). The sources and Einstein equations (\ref{eq:pertEq}) read accordingly:
\begin{equation}
\label{eq:SAxial_order_i}
\left( \pert{i}{S}_{\alpha\beta} \right)=
\left(\begin{array}{cccc}
    0 & 0 & 0 &  \pert{i}{S}_{03} \\
    0 & 0 & 0 &  \pert{i}{S}_{13} \\
    0 & 0 & 0 &  \pert{i}{S}_{23} \\
    \pert{i}{S}_{03} & \pert{i}{S}_{13} & \pert{i}{S}_{23} &  0 \\
\end{array}\right) \, ,
\end{equation}
with the components expanded according to
(\ref{eq:axialT_a3},\ref{eq:axialT_23}). In the axial case the sources
fulfill the following identity:
\begin{equation}
\label{eq:zero3}
\pert{i}{\Nsc}_{\ell\,\,3} := -\frac{1}{\fsc} \partial_t \pert{i}{S}_{\ell\,\,03} + \fsc \partial_r \pert{i}{S}_{\ell\,\,13} + \frac{2}{r}(2 \fsc - 1) \pert{i}{S}_{\ell\,\,13} - \frac{(\ell-1)(\ell+2)}{r^2} \pert{i}{S}_{\ell\,\,23} = 0\, .
\end{equation}
This is again obtained by taking the background divergence of
(\ref{eq:pertEq}): $\pert{i}{\Nsc}_{\ell\,\,3} = 0$ follows from the (one)
axial component of $\bar \nabla^{\mu} \pert{i}{E}_{\mu\nu} = 0$ (the (one)
axial component of $\bar \nabla^{\mu} \Delta_L \pert{i}{h}_{\mu\nu}$
vanishes identically thus the (one) axial component of $\bar
\nabla^{\mu} \pert{i}{S}_{\mu\nu}$ vanishes as well and (only) in that
sense the formula $\bar \nabla^{\mu} \pert{i}{S}_{\mu\nu}=0$ in \cite{dhs}
is correct for axial perturbations). Similarly to the polar case, the
gauge degrees of freedom enter $\pert{i}{E}_{\ell\,\,\mu\nu}$ only through
the source terms $\pert{i}{S}_{\ell\,\,\mu\nu}$ (and
$\pert{i}{S}_{\ell\,\,\mu\nu}$ depend on gauge functions ${}^{(j)}\eta_{\ell}$
with $j<i$). Thus we solve (\ref{eq:pertEq}) at order $i$ for
Regge-Wheeler variables $\pert{i}{f}_{\ell\,\,03}$, $\pert{i}{f}_{\ell\,\,13}$
and then recover aAdS gauge with a suitable gauge transformation (see
below). In the axial case the perturbative Einstein equations can be
easily integrated and the definition of the master scalar variable is
straightforward. We set
\begin{equation}
\label{eq:PhiAdef}
\pert{i}{f}_{\ell\,\,13} = \frac{r}{\fsc} \pert{i}{\Phi}^{\Asc}_{\ell}\, .
\end{equation} 
For $\ell \geq 2$ the three perturbative Einstein equations $\pert{i}{E}_{\ell\,\,03}$, $\pert{i}{E}_{\ell\,\,13}$ and $\pert{i}{E}_{\ell\,\,23}$ can be combined as follows (the $\ell=1$ case has to be treated separately, as $\pert{i}{E}_{1\,\,23} \equiv 0$). First
\begin{equation}
0 = -\frac{1}{\fsc} \partial_t \pert{i}{E}_{\ell\,\,03} + \fsc \partial_r \pert{i}{E}_{\ell\,\,13} + \frac{2}{r}\left(2 \fsc - 1\right) \pert{i}{E}_{\ell\,\,13} - \frac{(\ell-1)(\ell+2)}{r^2} \pert{i}{E}_{\ell\,\,23} 
\end{equation}
reduces to the identity (\ref{eq:zero3}). Second
\begin{equation}
\label{eq:NH_axial_wave}
0 = \frac{2 \fsc}{r} \left( \pert{i}{E}_{\ell\,\,13} - \partial_r \pert{i}{E}_{\ell\,\,23} \right) + \frac{4}{r^2} \, \pert{i}{E}_{\ell\,\,23} \equiv \tilde \Box_{\ell} \, \pert{i}{\Phi}^{\Asc}_{\ell} - \pert{i}{\tilde S}^{\Asc}_{\ell}\, , 
\end{equation}
where the $\tilde \Box_{\ell}$ operator was defined in (\ref{eq:H_scalar_wave}) and the axial source $\pert{i}{S}^{\Asc}_{\ell}$ reads: 
\begin{equation}
\label{eq:source_axial_wave}
\pert{i}{\tilde S}^{\Asc}_{\ell} = \frac{2 \fsc}{r} \left( \pert{i}{S}_{\ell\,\,13} - \partial_r \pert{i}{S}_{\ell\,\,23} \right) + \frac{4}{r^2} \, \pert{i}{S}_{\ell\,\,23} \, .
\end{equation}
Finally 
\begin{equation}
\label{eq:fT0}
2 \fsc \, \pert{i}{E}_{\ell\,\,23} = - \partial_t \pert{i}{f}_{\ell\,\,03} + \fsc \partial_r \left(\fsc \, \pert{i}{f}_{\ell\,\,13} \right) - 2 \fsc \, \pert{i}{S}_{\ell\,\,23} = 0 
\end{equation}
can be easily integrated in time for $\pert{i}{f}_{\ell\,\,03}$. 
Similarly to the polar case RW gauge (i.e. $\pert{i}{\eta}_{\ell} \equiv 0$ in (\ref{eq:h03}-\ref{eq:h23})) is not aAdS and the necessary condition for a gauge transformation to aAdS gauge to exist is $a_{\ell} \equiv 0$ in (\ref{eq:Box_asymptotics}), that is the correct boundary condition at infinity for master scalar variable for axial perturbations $\pert{i}{\Phi}^{\Asc}_{\ell}$ read
\begin{equation}
\label{eq:axialBC}
\pert{i}{\Phi}^{\Asc}_{\ell} = \frac{\pert{i}{b}_{\ell}}{r} + \mathcal{O}\left(\frac{1}{r^3}\right) \, .
\end{equation}
Once the boundary condition (\ref{eq:axialBC}) is satisfied the aAdS
gauge can be recovered with the axial gauge vector induced by
(cf. (\ref{eq:axialV_3}))
\begin{equation}
\label{eq:gaugeW}
\pert{i}{\eta}_{\ell} = \frac{\lsc^2}{3} \, \pert{i}{\Phi}^{\Asc}_{\ell} \, .
\end{equation}  

To summarize: to satisfy the set of perturbative Einstein equations
(\ref{eq:pertEq}) for axial type perturbations at axial symmetry, at
any nonlinear order, for angular momenta $\ell \geq 2$
(cf. (\ref{eq:axialT_a3},\ref{eq:axialT_23})), it is enough to solve
just one inhomogeneous wave equation for scalar master variable for
axial perturbations (\ref{eq:NH_axial_wave}) with the source term
given in (\ref{eq:source_axial_wave}), and then obtain the
Regge-Wheeler gauge invariant potentials $\pert{i}{f}_{\ell\,\,03}$,
$\pert{i}{f}_{\ell\,\,13}$ from (\ref{eq:PhiAdef}) and
(\ref{eq:fT0}). Then with the gauge transformation (\ref{eq:gaugeW})
the resulting perturbations $\pert{i}{h}_{\ell\,\,\mu\nu}$
(cf. (\ref{eq:h03}-\ref{eq:h23})) can be put in asymptotically AdS
form.

As was stressed above the special case $\ell=1$ needs a special treatment.
\subsection{the $\ell=1$ case for axial perturbations.} 
In the $\ell=1$ case the nontrivial equations are $0 =
\pert{i}{E}_{1\,\,13} = \pert{i}{E}_{1\,\,03}$ and either $\pert{i}{f}_{1\,\,03}$
or $\pert{i}{f}_{1\,\,13}$ becomes gauge degree of freedom (i.e. it can be
put to zero with a suitable choice of $\pert{i}{\eta}_{1}$).

If we decide to put $\pert{i}{f}_{1\,\,03}$ to zero then from
$\pert{i}{E}_{1\,\,13}=0$ we get
\begin{equation}
\partial_t \pert{i}{f}_{1\,\,13} = \pert{i}{f}^{\Asc \,\, res}_{1} + 2 \fsc \int^t \pert{i}{S}_{1\,\,13}\, dt' \, 
\end{equation}
where $\pert{i}{f}^{\Asc \,\, res}_{1} \equiv \pert{i}{f}^{\Asc \,\, res}_{1}(r)$ is set from $\pert{i}{E}_{1\,\,03}=0$:
\begin{equation}
\pert{i}{f}^{\Asc \,\, res}_{1} = \frac{2}{r^2} \int_0^r \frac{r'^2} {\fsc} \, \pert{i}{\bar S}_{1\,\,03}\, dr' \, ,  
\end{equation}
where $\pert{i}{\bar S}_{1\,\,03}$ is the time independent part of
$\pert{i}{S}_{1\,\,03}$ and we used (\ref{eq:zero3}).

If we decide to put $\pert{i}{f}_{1\,\,13}$ to zero then from
$\pert{i}{E}_{1\,\,13}=0$ we get
\begin{equation}
\pert{i}{f}_{1\,\,03} = \pert{i}{f}^{\Asc \,\, res}_{1} - 2 r^2 \int_{\infty}^r \frac{\fsc}{r'^2} \left(\int^t \pert{i}{S}_{1\,\,13}\, dt'\right) \, dr' \, , 
\end{equation}
where $\pert{i}{f}^{\Asc \,\, res}_{1} \equiv \pert{i}{f}^{\Asc \,\, res}_{1}(r)$ can be easily obtained from $\pert{i}{E}_{1\,\,03}=0$:
\begin{equation}
\partial_r \left[ \frac{1}{r^2} \partial_r \left(r \pert{i}{f}^{\Asc \,\, res}_{1} \right)\right] = - \frac{2}{r \, \fsc} \, \pert{i}{\bar S}_{1\,\,03} \, ,  
\end{equation}
where $\pert{i}{\bar S}_{1\,\,03}$ is the time independent part of
$\pert{i}{S}_{1\,\,03}$ and we used (\ref{eq:zero3}).

At linear order the $\ell=1$ part of the perturbation reduces to a
pure gauge, as expected (as ${}^{(1)}S_{\mu\nu} \equiv 0$).

\section{Application: perturbative construction of time-periodic solutions, preliminaries.}

After discussing our approach to (nonlinear) perturbations of AdS in
previous sections we apply it on some examples. In this
section we present some preliminary results on perturbative
construction of globally regular, time-periodic, aAdS solutions of
Einstein equations. Although for more systematic studies of TP
solutions it is necessary to go beyond the third order in perturbation
expansion (that we postpone for future studies), already the results
from the third order can provide some intuitions about TP solutions.

We start with gathering some spectral properties of gravitational
perturbations of AdS. Separating the time dependence in the
homogeneous wave equation (\ref{eq:H_scalar_wave}) $\tilde
\Box_{\ell}\Phi = 0$ in the form $\Phi(t,r) = e(r) \, \cos(\omega t)$
we get
\begin{equation}
\fsc \left[ - \frac{d}{dr} \left( \fsc \, \frac{d}{dr} e \right) + \frac{\ell(\ell+1)}{r^2} \, e \right] = \omega^2 e \,. 
\end{equation}
The frequencies $\omega$ are quantized by two boundary conditions:
regularity at $r=0$ and the required asymptotic behavior at infinity,
(\ref{eq:polarBC}) and (\ref{eq:axialBC}) for polar and axial modes
respectively, and the spectra of linear perturbations of AdS (AdS
eigenfrequencies) read
\begin{equation}
\label{eq:AdS_spectrum}
\lsc \omega^{\Psc}_{\ell,j} = 1 + \ell + 2 j \quad \mbox{and} \quad \lsc \omega^{\Asc}_{\ell,j} = 2 + \ell + 2 j \, ,
\end{equation}
where nonnegative integers $j$ are nodal numbers of the corresponding
eigenfunctions (AdS eigenmodes)
\begin{align}
\label{modes_polar}
e^{\Psc}_{\ell,j}(r) & = \mathcal{N}^{\Psc}_{\ell,j} \frac{\lsc r^{\ell + 1}} {\left(\lsc^2 + r^2\right)^{\frac{\ell + 1}{2}}} \, {}_2F_1\left(-j,1+\ell+j; 1/2; \lsc^2 / (\lsc^2 + r^2) \right) \, ,
\\
\label{modes_axial}
e^{\Asc}_{\ell,j}(r) & = \mathcal{N}^{\Asc}_{\ell,j} \frac{\lsc r^{\ell + 1}} {\left(\lsc^2 + r^2\right)^{\frac{\ell + 2}{2}}} \, {}_2F_1\left(-j,2+\ell+j;3/2;\lsc^2 / (\lsc^2 + r^2) \right) \, .
\end{align}
Thus all AdS eigenfrequencies are real and AdS is linearly stable. The
AdS eigenmodes with a given $\ell$ form a complete orthogonal set with respect to the scalar product
\begin{equation}
\label{eq:scalar_prod}
\left(u, \, v\right) \equiv \int_{0}^{\infty} \frac{u(r)v(r)}{1+r^2 / \lsc^2} dr \, ,
\end{equation}
namely 
\begin{align}
\left(e^{\Psc}_{\ell,j}, \, e^{\Psc}_{\ell,k}\right) &= \frac{\lsc^3 \, \pi \, j! \, \Gamma \left(\ell + j + 3/2 \right)}{2 (1 + \ell + 2j) \Gamma \left(j + 1/2 \right) \Gamma (1 + \ell + j)} \left(\mathcal{N}^{\Psc}_{\ell,j}\right)^2 \delta_{j\,k} \, ,
\\
\left(e^{\Asc}_{\ell,j}, \, e^{\Asc}_{\ell,k}\right) &=\frac{\lsc \, \pi \, j! \, \Gamma \left(\ell + j + 3/2\right)}{8 (2 + \ell + 2j) \Gamma \left(j + 3/2 \right) \Gamma (2 + \ell + j)} \left(\mathcal{N}^{\Asc}_{\ell,j}\right)^2 \delta_{j\,k} \, .
\end{align}
To ease the comparison between our results and those of \cite{ds} we
take $\mathcal{N}^{\Psc}_{\ell,j} = \mathcal{N}^{\Asc}_{\ell,j} = 1$
in the following. At any instant of time the master scalar variables
$\pert{i}{\Phi}^{\Psc}_{\ell}$ and $\pert{i}{\Phi}^{\Asc}_{\ell}$ can be
expanded in the bases of polar and axial eigenmodes
\begin{equation}
\pert{i}{\Phi}^{\Psc|\Asc}_{\ell}(t,r) = \sum_j \pert{i}{c}^{\Psc|\Asc}_{\ell \,\,j}(t) \, e^{\Psc|\Asc}_{\ell,j}(r) \, .
\end{equation}
Similarly the inhomogeneous equations (\ref{eq:NH_scalar_wave}) and
(\ref{eq:NH_axial_wave}) can be projected on the eigenmodes leading to
the forced harmonic oscillator equations for Fourier like coefficients
$\pert{i}{c}^{\Psc|\Asc}_{\ell \,\,j}$:
\begin{equation}
\pert{i}{\ddot c}^{\Psc|\Asc}_{\ell \,\,j} +  \left(\omega^{\Psc|\Asc}_{\ell,j}\right)^2 \pert{i}{c}^{\Psc|\Asc}_{\ell \,\,j} = \left( e^{\Psc|\Asc}_{\ell,j},\, \fsc\, \pert{i}{\tilde S}^{\Psc|\Asc}_{\ell} \right) \, .
\end{equation}
Since the general solution to forced harmonic oscillator equation 
\begin{equation}
\ddot c(t)+\omega_{0}^2\, c(t)=a \cos (\omega t) 
\end{equation}
reads
\begin{equation}
\label{eq:forced_oscilator}
c(t) = \frac{\dot c(0)}{\omega_0}\sin \left(\omega_0 t\right) + c(0) \cos \left(\omega_0 t\right) +
\begin{cases}
{\displaystyle \frac{a \left(\cos (\omega t )-\cos \left(\omega_0 t\right)\right)}{\omega _0^2-\omega ^2}\,, } & \omega_0 \neq \omega\,, 
\\ & \\
{\displaystyle \frac{a}{2 \omega _0} t \sin \left(\omega _0 t\right)\,,} & \omega_0 = \omega \, ,
\end{cases}
\end{equation}
if the projection $\left( e^{\Psc|\Asc}_{\ell,j},\, \fsc\, \pert{i}{\tilde S}^{\Psc|\Asc}_{\ell} \right)$ is resonant i.e. it contains
harmonic time dependence $\cos(\omega^{\Psc|\Asc}_{\ell,j} t)$ or
$\sin(\omega^{\Psc|\Asc}_{\ell,j} t)$ then a secular term of the form
$t \sin(\omega^{\Psc|\Asc}_{\ell,j} t)$ or $t
\cos(\omega^{\Psc|\Asc}_{\ell,j} t)$ appears in $\pert{i}{c}^{\Psc|\Asc}_{\ell \,\,j}$ and the naive perturbation expansion
breaks down. Resumming of all such possible secular terms gives rise
to the resonant system (cf. \cite{bbgll,cev1} in the case of a
massless scalar field at spherical symmetry). It may be also possible
for some particular first order solution
$\pert{1}{\Phi}^{\Psc|\Asc}_{\ell}(t,r)$ to be dressed at higher orders
by a suitable choice of frequency corrections and free integration
constants (cf. (\ref{eq:forced_oscilator})) in such a way, to remove
all resonant terms at higher orders and then such
$\pert{1}{\Phi}^{\Psc|\Asc}_{\ell}(t,r)$ (suitably dressed at higher
orders) give rise to TP solutions. For Einstein--AdS--selfgravitating
massless scalar field system any linear eigenmode can be extended
to form TP solution \cite{mrPRL} due to the absence of the so called
(+++) resonances (see \cite{cev1} for a rigorous theorem). In a recent
work \cite{ds} Dias and Santos contrasted gravitational sector of
perturbations with the scalar one by noticing that only in special
cases (listed in Sec.6 in \cite{ds}) linear eigenmodes admit a
nonlinear extension to a regular time-periodic solution. The reason is
that at the third order of the perturbation expansion around most
eigenmodes there appear resonant terms that, in contrast to
spherically symmetric scalar perturbations studied in \cite{br,mrPRL},
cannot be removed by a frequency correction. However, this is a purely
technical obstruction in constructing TP solutions, due to the
degeneracy of the spectrum and, when this degeneracy is properly taken
into account, one can construct perturbatively a TP solution
bifurcating from \emph{each} linear eigenfrequency as was pointed out
in \cite{r}. Here we complete \cite{r} with some more examples of TP
solutions bifurcating from degenerated eigenfrequencies. We aim at
constructing TP solution in the form
\begin{equation}
\label{eq:TP_general}
\Phi(t,r,\theta) = \pert{1}{\Phi}(t,r,\theta) \, \epsilon + \pert{2}{\Phi}(t,r,\theta) \, \epsilon^2 + \pert{3}{\Phi}(t,r,\theta) \, \epsilon^3 + \mathcal{O}\left( \epsilon^4 \right) 
\end{equation}
taking for the seed $\pert{1}{\Phi}$ a linear combination of \textit{all}
polar (axial) modes corresponding to a given eigenfrequency.

\paragraph{Polar modes with $\lsc \omega^{\Psc}=5$ as the seed.} 
If we start with
\begin{equation}
\label{eq:wP5}
\pert{1}{\Phi}(t,r,\theta) = \left(\eta \, e^{\Psc}_{2,1}(r) P_2(\cos \theta) + (1-\eta) \, e^{\Psc}_{4,0}(r) P_4(\cos \theta)\right) \cos\left((5 + \pert{2}{\omega} \, \epsilon^2)t / \lsc \right) \,
\end{equation}
at linear order \cite{r}, then at the third order we get two resonant terms
(for the modes $e^{\Psc}_{2,1}$ and $e^{\Psc}_{4,0}$) that can be
removed by a suitable choice of the frequency correction $\pert{2}{\omega}$
and the mixing parameter $\eta$ in \eqref{eq:wP5}. More precisely, the
resonant terms will be absent if $\pert{2}{\omega}$ and $\eta$ satisfy the
following system of equations:
\begin{align}
\label{eq:wP5l2}
-651980329\, \eta^3 +673396185\,\eta^2 -358711575\,\eta + 22494375 &= 49201152 \, \eta \,\pert{2}{\omega} \, ,
\\
\label{eq:wP5l4}
16847182891\, \eta^3 - 38330631185\, \eta^2 + 31825994625\, \eta - 10200766875 &= 4182097920 \, (1-\eta ) \,\pert{2}{\omega} \,.
\end{align}
This system has two real solutions: $(\eta, \pert{2}{\omega}) \approx
(0.1143,\,-1.900)$ and $(\eta, \pert{2}{\omega}) \approx (1.007, \,-6.487)$,
thus we expect two one parameter ($\epsilon)$ families of TP
solutions to bifurcate from the double eigenfrequency $\lsc
\omega^{\Psc}=5$. Note that setting $\eta=1$ in (\ref{eq:wP5l2}), we
get $\pert{2}{\omega} = -34397/5376$, while setting $\eta=0$ in
(\ref{eq:wP5l4}), we get $\pert{2}{\omega} = -52311625/21446656$, in agreement
with the values given in the Table~1 in \cite{ds}.

\paragraph{Polar modes with $\lsc \omega^{\Psc}=6$ as the seed.} 
If we start with
\begin{equation}
\label{eq:wP6}
\pert{1}{\Phi}(t,r,\theta) = \left(\eta \, e^{\Psc}_{3,1}(r) P_3(\cos \theta) + (1-\eta) \, e^{\Psc}_{5,0}(r) P_5(\cos \theta)\right) \cos\left((6 + \pert{2}{\omega} \, \epsilon^2)t / \lsc \right) \,
\end{equation}
at linear order, then at the third order the two potential resonances
(for the modes $e^{\Psc}_{3,1}$ and $e^{\Psc}_{5,0}$) are removed iff
the following system holds:
\begin{align}
\label{eq:wP6l3}
\frac{405 \left(11274706467 \eta^3 - 9177443205 \eta^2 + 4622768829 \eta - 285923771\right)}{48432676864} &= - \frac{315}{128} \eta \,\pert{2}{\omega} \, ,
\\
\label{eq:wP6l5}
\frac{405 \left(1192723542 \eta^3 - 2267107311 \eta^2 + 1606711764 \eta - 494724335\right)}{30820794368} &= \frac{693}{512} (1 - \eta) \,\pert{2}{\omega} \,.   
\end{align}
This system has two real solutions: $(\eta, \pert{2}{\omega}) \approx (0.1077,
\, -3.770)$ and $(\eta, \pert{2}{\omega}) \approx (1.022, \, -22.92)$, thus we
expect two one parameter families of TP solutions to bifurcate from the double eigenfrequency
$\lsc \omega^{\Psc}=6$.

\paragraph{Axial modes with $\lsc \omega^{\Asc}=6$ as the seed.} 
If we start with
\begin{equation}
\label{eq:wA6}
\pert{1}{\Phi}(t,r,\theta) = \left(\eta \, e^{\Asc}_{2,1}(r) P_2(\cos \theta) + (1-\eta) \, e^{\Asc}_{4,0}(r) P_4(\cos \theta)\right) \cos\left((6 + \pert{2}{\omega} \, \epsilon^2)t / \lsc \right) \,
\end{equation}
at linear order, then at the third order the two potential resonances
(for the modes $e^{\Asc}_{2,1}$ and $e^{\Asc}_{4,0}$) are removed iff
the following system holds:
\begin{align}
\label{eq:wA6l2}
\frac{239225693 \eta^3 - 398050275 \eta^2 + 217126035 \eta - 14644125}{9446621184} &= -\frac{35}{384} \eta  \pert{2}{\omega} \, ,
\\
\label{eq:wA6l4}
\frac{-4829869093 \eta^3 + 13136236344 \eta^2 - 12398007825 \eta + 4082879250}{267654266880} &= -\frac{63}{512} (1 - \eta) \pert{2}{\omega} \,.
\end{align}
This system has two real solutions: $(\eta, \pert{2}{\omega}) \approx
(0.1794, \, -0.08338)$ and $(\eta, \pert{2}{\omega}) \approx (1.008,
\, -0.05162)$, thus we expect two one-parameter families of TP
solutions to bifurcate from the double eigenfrequency $\lsc
\omega^{\Asc}=6$. Note that setting $\eta=1$ in (\ref{eq:wA6l2}) we
get $\pert{2}{\omega} = -19081/376320$, in agreement with the value
given in the second line in Table~2 in \cite{ds}.

\paragraph{Axial modes with $\lsc \omega^{\Asc}=7$ as the seed.} 
If we start with
\begin{equation}
\label{eq:wA7}
\pert{1}{\Phi}(t,r,\theta) = \left(\eta \, e^{\Asc}_{3,1}(r) P_3(\cos \theta) + (1-\eta) \, e^{\Asc}_{5,0}(r) P_5(\cos \theta)\right) \cos\left((7 + \pert{2}{\omega} \, \epsilon^2)t / \lsc \right) \,
\end{equation}
at linear order, then at the third order the two potential resonances
(for the modes $e^{\Asc}_{3,1}$ and $e^{\Asc}_{5,0}$) are removed iff
the following system holds:
\begin{align}
\label{eq:wA7l3}
\frac{1978248043512 \eta^3 - 2997546380577 \eta^2 + 1613025400770 \eta - 106858809617}{56956827992064} &= -\frac{21}{256} \eta \pert{2}{\omega} \, ,
\\
\label{eq:wA7l5}
\frac{3608791248537 \eta^3 - 9113490099567 \eta^2 + 8191761299127 \eta - 2669524922785}{144981016707072} &= \frac{231}{2048} (1 - \eta) \pert{2}{\omega} \,.
\end{align}
This system has two real solutions: $(\eta, \pert{2}{\omega}) \approx
(0.1670, \, -0.1130)$ and $(\eta, \pert{2}{\omega}) \approx (1.018, \,
-0.1085)$, thus we expect two one-parameter families of TP solutions
to bifurcate from the double eigenfrequency $\lsc \omega^{\Asc}=7$.

\paragraph{Some general remarks.} 
In general, for the eigenfrequency with geometric multiplicity $k$ we
need to take a linear combination of $k$ corresponding eigenmodes as
the seed, and to remove the resonances at the third order we have to
fulfill the system of $k$ equations that are cubic in the mixing
parameters $\eta_1,..,\eta_{k-1}$ and linear in $\pert{2}{\omega} \eta_i$
terms (with $1 \leq i \leq k-1$). Each \textit{real} root of this
system gives rise to a TP solution bifurcating from the given
eigenfrequency (the number of all (complex) roots is expected to grow
exponentially with $k$, as after eliminating $\pert{2}{\omega}$, we are left
with the system of $k-1$ equations that are quartic in
$\eta_1,..,\eta_{k-1}$). Interestingly, in all cases that we have
studied so far ($\lsc \, \omega^{\Psc} = 3,4,5,6,7,8$ and $\lsc \, \omega^{\Asc} =
4,5,6,7,8$ at axial symmetry), the number of bifurcating one-parameter
families of time-periodic solutions is \textit{equal} to the
multiplicity of the eigenfrequency. This intriguing coincidence
deserves further studies. Finally, we remark that in a parallel work
Maliborski constructed axially symmetric time-periodic solutions for
the cubic wave equation on the fixed AdS background \cite{m}, and in
that model there is no such coincidence.

\section{Conclusions.}
We have presented the formalism for nonlinear gravitational
perturbations of AdS spacetime and applied it to provide the evidence
for the existence and properties of globally regular, asymptotically
AdS, time-periodic solutions of vacuum Einstein equations. These time
periodic solutions bifurcate from the linear eigenfrequencies of AdS,
and the number of (one parameter families of) such solutions
bifurcating from a given eigenfrequency equals the multiplicity of
this eigenfrequency. This intriguing coincidence deserves further
studies. Already the results obtained at third order should provide
the initial guess for the numerical method of \cite{ds} good enough
for this method to converge and to provide an independent cross-check
of our results. The presented formalism should form the solid base for
the future construction of the resonant system for vacuum Einstein-AdS
equations, and the more systematic study of time-periodic solutions.

Although we were mainly motivated by the gravitational perturbations
of AdS, all the formulae in the paper that are not related to the
asymptotic structure of spacetime are general enough to encompass
gravitational perturbations of the other two maximally symmetric
vacuum solutions of Einstein equations, i.e. Minkowski and de Sitter
spacetimes, and with some modifications also spherically symmetric
vacuum solutions and as such will hopefully find application in the
broad subject of perturbations of such spacetimes. It would be also
very interesting to introduce matter in this context into the
presented formalism.

\noindent
\subsubsection*{Acknowledgements.} This research was supported in part by the Polish National Science Centre grant no. DEC-2012/06/A/ST2/00397. The author is indebted to Piotr Bizo\'n and Maciej Maliborski for many valuable discussions and to Piotr Bizo\'n for many valuable remarks on the early version of the manuscript. The author acknowledges the six months \textit{scientific associate} contract at CERN TH department that allowed him for the detailed studies of gravitational perturbations of AdS, and is particularly grateful to Luis \'Alvarez-Gaum\'e for providing propitious atmosphere for the research. A few days after the first version of this paper was released, there appeared a very interesting parallel work \cite{mfgf} on aAdS gravitational geons. In particular, in \cite{mfgf} the families of excited geons, bifurcating from degenerated linear frequencies of AdS, were numerically constructed. 

\end{widetext}


\begin{thebibliography}{10}

\bibitem{br}
P. Bizo\'n, A. Rostworowski, 
\textit{Weakly Turbulent Instability of Anti–de Sitter Spacetime}, 
Phys. Rev. Lett. \textbf{107}, 031102 (2011), \href{http://arxiv.org/abs/1104.3702}{\texttt{[arXiv:1104.3702]}}

\bibitem{mrPRL} 
M. Maliborski, A. Rostworowski, 
\textit{Time-Periodic Solutions in an Einstein AdS--Massless-Scalar-Field System}, 
Phys. Rev. Lett.  111, 051102 (2013), \href{http://arxiv.org/abs/1303.3186}{\texttt{[arXiv:1303.3186]}}

\bibitem{mr@bll} 
M. Maliborski, A. Rostworowski, 
\textit{A comment on ”Boson stars in AdS”}, 
\href{http://arxiv.org/abs/1307.2875}{\texttt{[arXiv:1307.2875]}}

\bibitem{bll_BS} A. Buchel, S.L. Liebling, L. Lehner, 
\textit{Boson Stars in AdS}, 
Phys. Rev. \textbf{D87}, 123006 (2013), \href{http://arxiv.org/abs/1304.4166}{\texttt{[arXiv:1304.4166]}}

\bibitem{dhms}
O.J.C. Dias, G.T. Horowitz, D. Marolf, J.E. Santos, 
\textit{On the Nonlinear Stability of Asymptotically Anti-de Sitter Solutions}, 
Class. Quant. Grav. \textbf{29}, 235019 (2012), \href{http://arxiv.org/abs/1208.5772}{\texttt{[arXiv:1208.5772]}}

\bibitem{mrPRD} 
M. Maliborski, A. Rostworowski, 
\textit{What drives AdS spacetime unstable?}, 
Phys. Rev. \textbf{89}, 124006 (2014), \href{http://arxiv.org/abs/1403.5434}{\texttt{[arXiv:1403.5434]}}

\bibitem{bbgll}
V. Balasubramanian, A. Buchel, S.R. Green, L. Lehner, S.L. Liebling, 
\textit{Holographic Thermalization, stability of AdS, and the Fermi-Pasta-Ulam-Tsingou paradox}, 
Phys. Rev. Lett. \textbf{113}, 071601 (2014), \href{http://arxiv.org/abs/1403.6471}{\texttt{[arXiv:1403.6471]}}

\bibitem{cev1}
B. Craps, O. Evnin, J. Vanhoof, 
\textit{Renormalization group, secular term resummation and AdS (in)stability}, 
JHEP \textbf{10} (2014), 48, \href{http://arxiv.org/abs/1407.6273}{\texttt{[arXiv:1407.6273]}}

\bibitem{cev2}
B. Craps, O. Evnin, J. Vanhoof, 
\textit{Renormalization, averaging, conservation laws and AdS (in)stability}, 
JHEP \textbf{01} (2015), 108, \href{http://arxiv.org/abs/1412.3249}{\texttt{[arXiv:1412.3249]}}

\bibitem{bmr}
P.~Bizo\'n, M.~Maliborski, A. Rostworowski, 
\textit{Resonant Dynamics and the Instability of Anti–de Sitter Spacetime}, 
Phys. Rev. Lett. \textbf{115}, 081103 (2015), \href{http://arxiv.org/abs/1506.03519}{\texttt{[arXiv:1506.03519]}}



\bibitem{hs}
G.T. Horowitz, J.E. Santos, 
\textit{Geons and the Instability of Anti-de Sitter Spacetime}, 
Surv. Differ. Geom. \textbf{20}, 321-335 (2015), \href{http://arxiv.org/abs/1408.5906}{\texttt{[arXiv:1408.5906]}}

\bibitem{dhs}
O.J.C. Dias, G.T. Horowitz J.E. Santos, 
\textit{Gravitational Turbulent Instability of Anti-de Sitter Space}, 
Class. Quant. Grav. \textbf{29}, 194002 (2012), \href{http://arxiv.org/abs/1109.1825}{\texttt{[arXiv:1109.1825]}}

\bibitem{ds}
O.J.C. Dias, J.E. Santos, 
\textit{AdS nonlinear instability: moving beyond spherical symmetry}, 
\href{http://arxiv.org/abs/1602.03890}{\texttt{[arXiv:1602.03890]}}

\bibitem{r}
A.~Rostworowski, 
\textit{Comment on "AdS nonlinear instability: moving beyond spherical symmetry"
}, 
\href{http://arxiv.org/abs/1612.00042}{\texttt{[arXiv:1612.00042]}}

\bibitem{Evnin}
O. Evnin, 
\textit{AdS perturbations, isometries, selection rules and the Higgs oscillator}, 
JHEP \textbf{01} (2016), 151, \href{http://arxiv.org/abs/1512.00349}{\texttt{[arXiv:1512.00349]}}

\bibitem{RW}
T.Regge and J.A. Wheeler, \textit{Stability of a Schwarzschild Singularity}, Phys.Rev \textbf{108}, 1063 (1957)

\bibitem{Nollert}
H.-P. Nollert, \textit{Quasinormal modes: the characteristic ‘sound’ of black holes and neutron stars}, Class. Quantum Grav. \textbf{16} R159–R216 (1999)

\bibitem{zerilli}
F.J. Zerilli, \textit{Effective potential for even-parity Rege-Wheeler Gravitational perturbation equations}, Phys. Rev. Lett. \textbf{24}, 737 (1970)

\bibitem{Mukohyama}
S. Mukohyama, \textit{Gauge-invariant gravitational perturbations of maximally symmetric spacetimes}, Phys. Rev. \textbf{62}, 084015 (2000)
\href{http://arxiv.org/abs/hep-th/0004067}{\texttt{[arXiv:hep-th/0004067]}}
 
\bibitem{BMMS}
M.~Bruni, S.~Matarrese, S.~Mollerach and S.~Sonego, \textit{Perturbations of spacetime: gauge transformations and gauge invariance at second order and beyond}, Class. Quantum Grav. \textbf{14} 2585–2606 (1997)
\href{http://arxiv.org/abs/gr-qc/9609040}{\texttt{[arXiv:gr-qc/9609040]}}

\bibitem{GP}
A. Garat and R.H. Price, \textit{Gauge invariant formalism for second order perturbations of Schwarzschild spacetimes}, Phys. Rev. \textbf{61}, 044006 (2000)
\href{http://arxiv.org/abs/gr-qc/9909005}{\texttt{[arXiv:gr-qc/9909005]}}

\bibitem{KI}
H. Kodama and A. Ishibashi, \textit{A Master Equation for Gravitational Perturbations of Maximally Symmetric Black Holes in Higher Dimensions}, Prog. Theor. Phys. \textbf{110}, 701 (2003)
\href{http://arxiv.org/abs/hep-th/0305147}{\texttt{[arXiv:hep-th/0305147]}}

\bibitem{BMGT}
D. Brizuela, J.M. Mart\'in-Garc\'ia and M. Tiglio, \textit{Complete gauge-invariant formalism for arbitrary second-order perturbations of a Schwarzschild black hole}, Phys. Rev. \textbf{80}, 024021 (2009)
\href{http://arxiv.org/abs/0903.1134}{\texttt{[arXiv:0903.1134]}}

\bibitem{Moncrief}
V. Moncrief, \textit{Gravitational Perturbations of Spherically Symmetric Systems. I. The Exterior Problem}, Ann. Phys. \textbf{88}, 323 (1974)

\bibitem{HenneauxTeitelboim}
M. Henneaux and C. Teitelboim, \textit{Asymptotically Anti-de Sitter Spaces}, Commun. Math. Phys. \textbf{98}, 391-424 (1985)

\bibitem{BantilanPretoriusGubser}
H. Bantilan, F. Pretorius, and S. Gubser, \textit{Simulation of asymptotically $AdS_5$ spacetimes with a generalized harmonic evolution scheme}, Phys. Rev. \textbf{D85}, 084038 (2012), \href{http://arxiv.org/abs/1201.2132}{\texttt{[arXiv:1201.2132]}}

\bibitem{m}
M.~Maliborski, 
\textit{Cubic wave equation on AdS with axial symmetry – a toy model of geons}, 
to appear

\bibitem{mfgf}
G.~Martinon, G.~Fodor, P.~Grandcl\'ement and P. Forg\'acs, \textit{Gravitational geons in asymptotically anti-de Sitter spacetimes}, \href{http://arxiv.org/abs/1701.09100}{\texttt{[arXiv:1701.09100]}} 



    

















\end{thebibliography}
\end{document}